\documentclass[12pt,a4paper]{article}
\pdfoutput=1
\usepackage[english]{babel}
\usepackage[latin1]{inputenc}
\usepackage[T1]{fontenc}
\usepackage{amsfonts,amsbsy,bm,euscript,mathrsfs}
\usepackage{amssymb,faktor,slashed}
\usepackage{color}
\usepackage[tbtags]{amsmath}
\usepackage[bookmarks=true,colorlinks=true,linkcolor=black,citecolor=black,urlcolor=black,bookmarksnumbered,linktocpage=true]{hyperref}
\usepackage{graphicx}
\usepackage{chngcntr}
\usepackage{makecell}
\usepackage{mathtools}

\usepackage[a4paper,text={170mm,257mm},centering]{geometry}

\numberwithin{equation}{section}

\renewcommand{\arraystretch}{1.3}

\makeatletter
\renewcommand\section{\@startsection {section}{1}{\z@}%
{-3.5ex \@plus -1ex \@minus -.2ex}%
{2.3ex \@plus.2ex}%
{\normalfont\large\bfseries}}
\renewcommand\subsection{\@startsection{subsection}{2}{\z@}%
{-3.25ex\@plus -1ex \@minus -.2ex}%
{1.5ex \@plus .2ex}%
{\normalfont\normalsize\bfseries}}
\makeatother

\expandafter\def\expandafter\bfseries\expandafter{\bfseries\ifmmode\else\boldmath\fi}
\expandafter\def\expandafter\mdseries\expandafter{\mdseries\ifmmode\else\unboldmath\fi}
\expandafter\def\expandafter\normalfont\expandafter{\normalfont\ifmmode\else\unboldmath\fi}

\providecommand{\href}[2]{#2}
\newcommand{\arxivlink}[1]{\href{http://arxiv.org/abs/#1}{[arXiv:#1]}}
\newcommand{\doilink}[2]{\href{http://doi.org/#2}{#1}}

\newcommand{\mathsym}[1]{{}}
\def\id{\protect{{1 \kern-.28em{\rm l}}}}
\def\be{\begin{eqnarray}}
\def\ee{\end{eqnarray}}
\def\bi{\bibitem}

\def\ha{\tfrac{1}{2}}
\def\td{\tilde}
\def\ci{\cite}

\def\a{\alpha}
\def\b{\beta}

\def\g{\gamma}

\def\Tr{{\rm Tr}}

\def\l {\lambda}

\def\const{{\rm const}}

\def\O{{\mathcal O}}
\def\m{\mu}

\def\foot{\footnote}
\newcommand{\rf}[1]{(\ref{#1})}

\def\bw{{\rm w}}

\def\no{\nonumber}

\def\la{\label}
\def\l{\lambda}

\def\p{\phi}
\def\r{\rho}

\def\varpi{{\rm w}}

\def\t{\tau}

\def\del{\partial}
\def\s{\sigma}

\def\n{\nu}

\def\ed{\end{document}}
\newcommand{\mc}{\mathcal }
\def\iffa{\iffalse}

\def\d{\delta}

\def\L{\mathcal{L} }

\def\sm{$\sigma$-model}
\def\sms{$\sigma$-models}

\def\Ad{\text{Ad}}

\def\Lie{\operatorname{Lie}}

\def\ddt{\frac{d}{dt}}

\def\M{{\cal M}}

\def\ed{\end{document}}

\def\cgg{c_{_G}}
\def\cg{c\,}

\def\e{\varepsilon}

\def\GG{{\rm G}}

\def\g{\gamma}

\def\hz{z}
\def\hhz{z} 
\def\wcL{\widehat{\L}}
\def\wL{\widehat{L}}

\def\bw{\a}

\def\cR{{\mathcal R}}
\def\sms{$\s$-models\ }
\def\sm{$\s$-model\ }
\def\wLL{\widehat{\L}}

\newcommand\ei[1]{\mathbf{\underline{#1}}}
\def \FF {h}
\def\GG{{\rm G}}
\begin{document}

\ \hfill{\small Imperial-TP-NL-2020-01 }

\vspace{0.5cm}


\vspace{1.5cm}

\begin{center}

{\Large\bf Sigma models with local couplings:\\
\vspace{0.3cm}
a new integrability--RG flow connection }

\vspace{1.5cm}
{
Ben Hoare$^{a,}$\footnote{\ bhoare@ethz.ch}, \
Nat Levine$^{b,}$\footnote{\ n.levine17@imperial.ac.uk} and
Arkady A. Tseytlin$^{b,}$\footnote{\ Also at the Institute of Theoretical and Mathematical Physics, MSU and Lebedev Institute, Moscow.
\\\hspace*{15pt} \ tseytlin@imperial.ac.uk}
}

\vspace{0.8cm}

{
\em \vspace{0.15cm}
$^{a}$Institut f\"ur Theoretische Physik, ETH Z\"urich,\\
\vspace{0.05cm}
Wolfgang-Pauli-Strasse 27, 8093 Z\"urich, Switzerland.
\\
\vspace{0.15cm}
$^{b}$Blackett Laboratory, Imperial College, London SW7 2AZ, U.K.
}

\end{center}

\vspace{0.5cm}

\begin{abstract}
We consider several classes of $\sigma$-models (on groups and symmetric spaces, $\eta$-models, $\lambda$-models) with local couplings that may depend on the 2d coordinates, e.g.\ on time $\tau$. We observe that (i) starting with a classically integrable 2d $\sigma$-model, (ii) formally promoting its couplings $h_\alpha$ to functions $h_\alpha(\tau)$ of 2d time, and (iii) demanding that the resulting time-dependent model also admits a Lax connection implies that $h_\alpha(\tau)$ must solve the 1-loop RG equations of the original theory with $\tau$ interpreted as RG time. This provides a novel example of an `integrability--RG flow' connection. The existence of a Lax connection suggests that these time-dependent $\sigma$-models may themselves be understood as integrable. We investigate this question by studying the possibility of constructing non-local and local conserved charges. Such $\sigma$-models with $D$-dimensional target space and time-dependent couplings subject to the RG flow naturally appear in string theory upon fixing the light-cone gauge in a $(D+2)$-dimensional conformal $\sigma$-model with a metric admitting a covariantly constant null Killing vector and a dilaton linear in the null coordinate.
\end{abstract}

\newpage

\tableofcontents

\setcounter{footnote}{0}
\setcounter{section}{0}

\section{Introduction}

There is a remarkable link between the (classical)
integrability of 2d $\s$-models and their stability under RG flow, i.e.\ their
renormalizability with finitely many running couplings (see, e.g., \cite{Fateev:2019xuq,HLT} and refs.\ there).
For example, the classically integrable $\eta$-deformed \cite{Klimcik:2002zj,Delduc:2013fga} and $\l$-deformed \cite{Sfetsos:2013wia,Hollowood:2014rla} models
are stable under the 1-loop RG flow with only the overall scale and the deformation parameter running
\cite{Valent:2009nv,Sfetsos:2014jfa,Appadu:2015nfa}.
However, it
remains rather mysterious how the (classical) integrability translates into simple
(quantum) RG behaviour.

In this paper we present a new and different link between
classical integrability and the RG flow.
Consider a familiar example of an integrable $\s$-model, the 2d
principal chiral model (PCM). Suppose we
formally replace its overall coupling $h$
by a function of 2d time $\t$,
obtaining
the
Lagrangian\foot{\la{fo1}We use 2d Minkowski space with metric
$\eta_{ab}=(-1,1)$ and
coordinates $(x^0,x^1) = (\t,\s)$. The
2d light-cone (l.c.) coordinates are $\xi^\pm = \ha (\t\pm \s)$ and the corresponding l.c.\ derivatives are $\del_\pm = \del_\t \pm \del_\s$. We follow the notation of \cite{HLT}, i.e.\ the current $J_a$ is given by
$J_a= g^{-1} \del_a g, \ g \in G$ and $ \Tr[J_+ J_-]= J^n_+ J^n_- = - 2\del_+ x^n \del_- x^n + ...$
for $g= e^{ i \sqrt 2 x^n T_n} $.
The action is defined as $S=\frac{1}{4 \pi} \int d \tau d \s \, \L$ and we set $\a'=1$.
Let us note that the model \rf{1} is not equivalent to the standard PCM
in a particular curved 2d metric: since any 2d \sm is classically Weyl invariant,
there is always a choice of coordinates (``conformal gauge'' in a string context)
in which its action takes the standard flat-space form.
}
\be\la{1}
\wcL = -\ha h(\t) \Tr[J_+ J_-] \ .
\ee
The resulting time-dependent theory is not Lorentz invariant but one may wonder if
some notion of classical integrability still applies even for a non-constant $h(\t)$.
As we shall see below,
this is indeed the case {\it provided} the function $h(\t)$ is special:
the same as the solution of the 1-loop RG equation of the standard PCM,
i.e.\ $h(\t) \sim \t$.

More generally, starting with a Lorentz-invariant
2d theory
that is classically integrable (i.e.\ admitting a Lax connection),
one may formally promote
its coupling constants $h_\a$ to functions $h_\a (\t)$ depending on 2d time.
We address the question of which functions $h_\a (\t)$
allow the integrability to survive in the resulting time-dependent theory, in the sense that the resulting equations of motion still
admit a Lax representation.
We shall suggest a certain natural ansatz for the corresponding Lax connection
generalizing that of the time-independent theory and
will find that the unique functions $h_\a (\t)$ preserving the integrability under this ansatz are the ones solving the 1-loop RG equations of the original theory,
\be
\del_\t h_\a = \b_\a(h) \ , \la{1lr}
\ee
with
2d time $\t$ playing the role of
RG time $t=\log \mu$.
We observe this new `integrability--RG flow' connection
in a number of non-trivial examples but will not give a general proof of it.

One may wonder how such
time-dependent models (where, e.g.,
energy is no longer conserved)
can be integrable. In fact, integrable examples of time-dependent 1d
mechanical models
are known (see, e.g., \cite{t-dep-1d} and refs. there).
Our present interest in such 2d models is motivated by the desire to find new solvable examples
of
conformal $\s$-models representing consistent string backgrounds with Minkowski signature.
For example, consider a conformal \sm with the $(D+2)$-dimensional
target space metric
$ds^2 =G_{\m\n} (x) dx^\m dx^\n = -2 dudv + K(u,x) du^2 + dx^i dx^i$ \ ($x^\m=(u,v,x^i)$, \ $i=1, ..., D$),
admitting a covariantly constant null Killing vector.
In the light-cone (l.c.) gauge $u=\tau$ one then gets
a 2d model with a time-dependent potential $K(\t,x)$, which is formally solvable in some cases if
$K$ is quadratic in $x^i$ (see, e.g., \cite{Papadopoulos:2002bg} and refs. there).\foot{Another solvable example obtained as a certain Yang-Baxter deformation of flat 4d Minkowski space was found in \cite{yoshida}.
Examples with $K$ not depending on $\t$ corresponding to
integrable l.c.-gauge theories (sine-Gordon, Liouville,
etc.) were discussed, e.g., in
\cite{Maldacena:2002fy,Russo:2002qj,Bakas:2002kt}. }

A remarkable class of string \sms with the
metric
$ds^2 = - 2 dud v + G_{ij}(u,x) d x^i d x^j$,
also admitting a
covariantly constant
null Killing vector,
but
with a {\it curved} transverse part
and also
a non-trivial dilaton,
was considered in \cite{Tseytlin:1992pq,Tseytlin:1992ee}. The corresponding classical string Lagrangian is
\begin{align}
\L =- 2 \del_+ u \del_- v + G_{ij}(u,x) \del_+ x^i \del_- x^j
\ , \la{csm}
\end{align}
and the dilaton term is $\L_\p= R^{(2)} \phi(v,u,x)$.
The central observation is that
this model is Weyl-invariant, i.e.\ $\bar \b_{\m\n} = R_{\m\n} + 2 D_\m D_\n\p + ...=0$,
provided $\phi$ is linear in the null coordinate $v$
and $G_{ij}(u,x)$ depends on $u$ according to the RG equation of the time-independent `transverse' $\s$-model
$\L=G_{ij}(x) \del_+ x^i \del_- x^j $,\foot{Let us note that the most general metric with a covariantly constant
null Killing metric has the Brinkmann-Walker form
$ds^2 = -2 dudv + K(u,x) du^2 + A_i(u,x) du dx^i + G_{ij} (u,x) dx^i dx^j$ where $K$ and $A_i$
terms may be absorbed into the $G_{ij}(u,x)$ term by a coordinate transformation
once one relaxes the assumption that $G_{ij}$ is flat.
The model \rf{csm} admits a natural generalization to the case of non-zero `transverse'
antisymmetric tensor coupling $B_{ij}(u,x)$: the resulting $(D+2)$-dimensional $\s$-model is Weyl invariant
provided $\del_u(G_{ij} + B_{ij}) = \bar{\beta}_{ij}(G,B)$
where $ \bar{\beta}_{ij}$ is the corresponding Weyl anomaly coefficient
(see
\cite{Tseytlin:1992pq,Tseytlin:1992ee} for details). In the first-order equation \rf{bd} we ignore an additional
diffeomorphism term $D_{(i}W_{j)}$ that appears only at higher loop orders.}
\begin{align}
& \del_u G_{ij} = \bar{\beta}_{ij} (G)= \beta_{ij} + 2 D_i D_j \bar{\phi}
\ , \qquad \quad \beta_{ij} = R_{ij} + \ldots
\ , \la{bd} \\
&\phi = v + \bar{\phi}(u,x) \ . \la{dilv}
\end{align}
The `first-order' equation \rf{bd} comes from the contribution of the connection in the $D_\m D_\n\p$ term
in the Weyl-invariance condition due to the null Killing form of the metric and the $v$-term in the dilaton.\foot{The exact
RG evolution equation \rf{bd} in the l.c.\ direction $u$
may be compared to an approximate RG equation
(see, e.g., \cite{Schmidhuber:1994bv}) appearing in the
case of a `cosmological' metric
and a time-dependent dilaton which provides a `friction' term in the string
generalization of the Einstein equations.
}

An example of such a conformal background with a sphere $S^D$ as the transverse space is \,
$ds^2 =- 2 du dv + u \, d\Omega^2_D, \ \ \p= v + \frac{1}{8} D \log u $ (to 1-loop order and after a rescaling of $u$ and $v$).
If the transverse part is an $\mathcal{N}=(2,2)$ supersymmetric $\s$-model on an Einstein-K\"{a}hler manifold, the $\b$-function
is given just by the 1-loop term and thus such a solution may be argued to be exact \cite{Tseytlin:1992ee}.

Starting with the classical string model \rf{csm} in conformal gauge and fixing the residual conformal symmetry by the l.c.\ gauge condition $u=\t$, we get a time-dependent
theory for the transverse coordinates\foot{Here
we ignore
the `diffeomorphism vector'
term that will also contribute to the time dependence of the metric; however,
this term can be eliminated or modified as desired by an appropriate field redefinition
in \rf{lct}
depending on time, $x^i \to y^i(x,\t)$,
see also footnote \ref{777}.}
\be
\L_{l.c.} = G_{ij}(\t ,x) \del_+ x^i \del_- x^j \ , \qquad\qquad
\del_\t G_{ij} = \beta_{ij}(G) \ . \la{lct}
\ee
If we now assume that the `transverse' $\s$-model
$\L=
G_{ij}(x) \del_+ x^i \del_- x^j$ is classically
integrable, then the fact that $G_{ij}(\t ,x)$ in \rf{lct} is subject to the RG equation
suggests, in view of the above `integrability--RG flow' connection,
that the l.c.\ gauge theory \rf{lct} is classically integrable
(admits a flat Lax connection)
and thus may potentially be solvable.

\medskip

Let us note that a special class of $\s$-models \rf{csm} with $G_{ij}(u,x) = u\, \GG_{ij}(x) $ may be viewed as
a gauge-fixed version
of a dilaton gravity (with metric $g_{ab}$) coupled to a $\s$-model \ci{Tseytlin:1992pq} with $\GG_{ij}(x)$ as the target space metric
\be \la{dig}
\L =\sqrt {-g} \, u \big[R^{(2)} + \GG_{ij}(x) \del^a x^i \del_a x^j \big] \ .
\ee
Here $u$ appears as the `dilaton' factor and $v$ plays the role of the conformal factor of the metric
upon fixing the conformal gauge $ds^2= g_{ab} d\s^a d \s^b = e^{-2v } \eta_{ab}d\s^a d \s^b , \ \sqrt {-g} \, R^{(2)} = 2 \del_+ \del_- v$. Particular 2d models of the type \rf{dig}, with $\GG_{ij}(x)$
corresponding to a specific symmetric space,
appear upon dimensional reduction (in 2 Killing vector directions) of 4d Einstein gravity
\ci{Belinsky:1971nt,Breitenlohner:1986um,Nicolai:1991tt}.
After solving $\del_+ \del_- u=0$ (the equation of motion of the 2d conformal factor $v$) to obtain $u(\t,\s) = f^+ (\xi^+) + f^- (\xi^-)$,
the resulting `local' $\s$-model,
\be \la{ff}
\L = u(\t, \s)\, \GG_{ij}(x) \del_+ x^i \del_- x^j \ , \qquad \quad u(\t,\s) = f^+ (\xi^+) + f^- (\xi^-) \ , \ee
was shown, in some special cases
\ci{Belinsky:1971nt} and for general symmetric spaces
\ci{Breitenlohner:1986um} (see also \ci{Nicolai:1991tt}), to be integrable in the sense of admitting a Lax
pair.
The spectral parameter is now replaced by a function of $\t,\s$ expressed in terms of $f^+ (\xi^+) $, $
f^- (\xi^-)$ and an integration constant $w$ (which may be viewed as a new spectral parameter --
see below).

Here will go beyond this related earlier work in two important ways:
(i) showing that there are other integrable local $\s$-models (e.g., related to integrable deformations of symmetric space $\s$-models)
for which the dependence of $G_{ij} (u,x)$ on $ u(\t,\s)$ is not simply through an overall factor as in \rf{ff};
(ii) establishing the link between the dependence on $u$ required for
integrability and the RG evolution of couplings in the original $\s$-model defined by the metric $G_{ij} (u,x)\big|_{u=\const}$.

\medskip

This paper is organized as follows.
In section \ref{sec2} we will show that, starting from a classically integrable $\s$-model
admitting a Lax pair, it is possible to find a Lax pair for the corresponding
time-dependent theory obtained from the original one by replacing the couplings
by functions of $\t$ satisfying the 1-loop RG equations. We will explicitly discuss the case of the PCM but a similar statement is
also true for several classes of $\s$-models (PCM with WZ term, coset model,
$\eta$-model, $\l$-models) presented in Table \ref{tab1}.

In section \ref{s3} we will demonstrate that the converse is also true: demanding the existence of a Lax pair for the time-dependent model implies that the couplings must solve the 1-loop RG equations
of the original theory. Details of the derivation of the RG flow will be presented in Appendix \ref{A}.

In section \ref{s4} we will discuss how the monodromy matrix associated with
the Lax connection can be used to construct, at least in principle, non-local
conserved charges for time-dependent models. We will concentrate on the PCM example
with comments on other models relegated to Appendix \ref{om}.
To shed light on the notion of integrability in the time-dependent case we
will also consider consistent 1d
reductions of some simple 2d $\s$-models demonstrating their solvability.
In Appendix \ref{C} we will further discuss a particular linear 1d model which is a special case
of a time-dependent harmonic oscillator.

In section \ref{s5} we
consider the PCM
with a local ($\s$-dependent)
coupling, which, like the time-dependent model, admits a Lax connection.
We work in the Hamiltonian formalism
to
investigate the existence of an infinite tower of local conserved charges in involution.
A naive generalization of the standard method used for constant couplings does not yield such charges.
However, in the case of the `chiral' model, i.e.\ with the coupling
depending
on the l.c.\ variable $\xi^-=\ha(\t-\s)$,
we find that with a slight modification it can be made to work.
In Appendix \ref{chiral} we show that such `chiral'
models correspondingly admit local holomorphic higher-spin currents.

Section \ref{s6} contains a discussion of the results and possible extensions.
In particular, we will mention that the relation between integrability of the time-dependent
theory and the RG flow should extend to models with potentials, illustrating this
on the example of the sine-Gordon model (see also Appendix \ref{sing}).

\section{Lax connection for time-dependent generalization of integrable \texorpdfstring{$\s$}{σ}-models}\la{sec2}

Below we shall discuss how,
starting from a classically integrable model (with Lagrangian $\L$) admitting a Lax pair $L_\pm$,
it is possible to find also a Lax pair $ \widehat{L}_\pm $ for the corresponding
time-dependent theory $\widehat{\L}$ obtained from the original one
by replacing the couplings $h_\a$ by functions $ h_\a (\t)$ depending on $\t$
according to the 1-loop RG equation \rf{1lr},\foot{Here and below for simplicity
we are only indicating the dependence of the Lagrangian on the couplings, suppressing dependence on fields.}
\begin{align}
&h_\a \to h_\a (\t) \ , \qquad \qquad \tfrac{d}{d\t} h_\a (\t) = \b_\a(h(\t)) \la{tins} \ ,\\
&\L(h_\a)\ \to\ \widehat{\L} = \L(h_\a (\t)) \ . \la{lcg}
\end{align}
The Lax connection must be modified to account for `extra' terms in the equations of motion following from \rf{lcg} that are proportional to the derivatives $\del_\t h_\a$ of the couplings.
We propose a very simple ansatz for this modification: the extra terms can be effectively `absorbed'
into the spectral parameter $z$
with the structure
of the Lax connection remaining the same, i.e.
\begin{align}
&L_\pm(h_\a, z) \to \widehat{L}_{\pm} = L_\pm (h_\a(\t), \hhz(w; \t,\s) )\ , \la{lac2}\\
&z \to \hhz = \hhz(w;\t,\s) \ , \ \ \ \qquad h_\a \to h_\a (\t) \ .
\la{lac1}
\end{align}
While $z$ was regarded as constant
in the original Lax connection,
it is now promoted to a particular function $\hhz(w;\t,\s)$ of the 2d coordinates.
Here $w$ is an arbitrary (complex) constant playing the role of a new spectral parameter.
As we shall see below, the explicit form of the function $\hhz(w;\t,\s)$ will be model-dependent.

Let us note that if the original theory is a \sm on flat 2d Minkowski space,
which
is classically invariant under the 2d conformal transformations $\xi^\pm \to f^{\pm}(\xi^\pm)$ (cf.\ footnote \ref{fo1}),
then the Lagrangian of the corresponding time-dependent theory \rf{lcg} changes
under such transformations as
\be
\widehat{\L}= \L(h_\a (\t))\ \to\ \L( h_\a(\t')) \ , \qquad \ \ \ \ \ \t = \xi^+ + \xi^-\ \to\ \t'=
f^+ (\xi^+)+ f^- (\xi^-) \ . \la{nlc}
\ee
In the case of the
string
\sm \rf{csm},\rf{lct},
this is equivalent to replacing the l.c.\ gauge $u=\t$ with a
more general one given by the solution $u = f^+ (\xi^+)+ f^- (\xi^-)$ of the
equation of motion $\del_+ \del_- u = 0$.
One special choice is $u=\xi^-$ for which
we get
$
\wcL_{\rm ch} = \L(h_\a (\xi^-))
$.
We may formally consider this theory,
which only fixes one `half' of the conformal reparametrizations, as an extreme limit $(f^+(\xi^+),f^-(\xi^-)) =(0, \xi^-)$ of \rf{nlc}.
We shall discuss such chiral theories further
in section \ref{s5} and Appendix \ref{chiral}, showing that they admit higher-spin conserved charges and holomorphic higher-spin local currents.

To provide evidence for the above proposal, we have considered
six examples of integrable $\s$-models with the results summarized in Table \ref{tab1} below.
There, for each model, we give the original Lagrangian, original Lax pair, 1-loop RG equation and its solution, and the expression for the function $\hhz(w;\t,\s)$ that generalizes the original spectral parameter in such a way
that \rf{lac2} gives a Lax pair for the time-dependent theory \rf{lcg}.\foot{\la{777}The classically integrable $\s$-models considered in this paper are 1-loop renormalizable, i.e.\
only finitely many couplings run under the 1-loop RG flow assuming a particular choice of
`renormalizable'
diffeomorphism vector. This means that the insertion of time dependence according to the RG flow in \rf{tins},\rf{lcg}
will only apply to a finite set of couplings. As discussed
in the Introduction,
a time-dependent $\s$-model
arises in the light-cone gauge from a conformal string $\s$-model.
There the time dependence is governed by the RG flow with
a specific diffeomorphism vector (generally different from the renormalizable one)
following from the $(D+2)$-dimensional dilaton solving the Weyl-invariance condition.
However,
these two patterns of time dependence
(according to the RG flow with different diffeomorphism vectors)
are related
by a time-dependent field redefinition in \rf{lcg}; such a redefinition preserves the existence of a Lax representation.}

All of our examples are built on either a group $G$ or a symmetric space $G/H$. We use the notation $J_\pm = g^{-1} \del_\pm g$, where $g\in G$ is scalar field.
PCM$_k$ stands for the principal chiral model plus a Wess-Zumino ($B$-field)
term with coefficient $k$
(the special case $h=\pm k$ is the conformal WZW model).
The group space $\eta$-model is an integrable deformation of the PCM
while the group space and symmetric space $\l$-models are integrable
deformations of the WZW and gauged WZW models
(we use the same definitions as in \cite{HLT}).

\begin{table}
{
\renewcommand{\arraystretch}{1.5}
\begin{tabular}{l || l | l | l }
& PCM& PCM$_k$ \\ \hline\hline
Lagrangian& $\L =-\ha h \Tr[J_+ J_-]$ & $\L =-\ha h \Tr[J_+ J_-]+ k\, \L_{\rm WZ}$\\ \hline

Lax connection & $L_\pm =\ha (1+z^{\pm 1})J_\pm$ & $L_\pm =\ha (1 + z^{\pm 1}) (1\pm \frac{k}{h})J_\pm$ \\ \hline

\makecell[l]{1-loop RG equation \\
and its solution}& $\begin{matrix*}[l]\ddt h = \cg \\ h(t) =\cg t \end{matrix*}$ & $\begin{matrix*}[l]\ddt h = \cg(1-\frac{k^2}{h^2}), \ \ \ \ddt k = 0 \\ h(t) -k \tanh^{-1}{\big(\frac{h(t)}{k}\big)} = \cg t \end{matrix*}$\\ \hline

\makecell[l]{Spectral parameter\\function }& $\hz=
\sqrt{\frac{w+\s-\t}{w+\s+\t}}$ & $\begin{matrix*}[l] \cg (w+\s-\t) + 2zh(\t)\, \frac{k+\hz(k+h(\t))}{k(1+\hz)^2+h(\t)(\hz^2-1)} \\ \quad \quad \qquad + k \log{\big(\frac{(1+\hz)(k-h(\t))}{(k-h(\t))+\hz(k+h(\t))}\big)} = 0 \end{matrix*}$
\vspace{0.5cm}
\end{tabular}

\begin{tabular}{l || l | l | l}
& Group space $\eta$-model & Group space $\l$-model \\ \hline\hline
Lagrangian& $\L =-\ha h \Tr [ J_+ \frac{1}{1-\eta \cR} J_- ] $ & $\begin{matrix*}[l]\L =k\big[ \L_{G/G}(g,A)\\
\qquad \qquad - (\l^{-1}-1) \Tr( A_+ A_-)\big] \end{matrix*}$\\ \hline

Lax connection & $\begin{matrix*}[l] L_\pm = \ha (1+z^{\pm 1}) C_\pm \\ C_\pm = - (1+\eta^2)\Ad_g\frac{1}{1\pm \eta \cR} J_\pm \end{matrix*}$ & $ \begin{matrix*}[l] L_\pm = \ha (1+z^{\pm 1}) \frac{2}{1+\l} A_\pm \\
A_\pm \text{ takes on-shell value}\end{matrix*}$ \\ \hline

\makecell[l]{1-loop RG equation \\
and its solution}
& $\begin{matrix*}[l] \ddt h = \cg (1+\eta^2)^2 ,\ \ \ddt (\eta h^{-1}) = 0\\
\nu \equiv \eta(t) h(t)^{-1} = \text{const} \\
{\tan}^{-1}\, {\eta(t)} + \frac{\eta(t)}{1+\eta(t)^2} = 2\cg\nu t
\end{matrix*}$
& $\begin{matrix*}[l]
\ddt \l = \frac{2\cg}{k}\frac{\l^2}{(1+\l)^2}, \ \ddt k =0, \\
2\log{\l(t)} + \l(t) - \l(t)^{-1} = \frac{2\cg}{k} t
\end{matrix*}$
\\ \hline

\makecell[l]{Spectral parameter \\
function}
& $\begin{matrix*}[l]
2\cg \nu (w+\s) - \frac{1-\hz}{1+\hz} \frac{\eta(\t)}{1+\eta(\t)^2} \\
\qquad \qquad + \tan^{-1}{ \big(\frac{1}{\eta(\t)} \frac{1-\hz}{1+\hz} \big)} = 0
\end{matrix*}$ &
$\begin{matrix*}[l]
\frac{2\cg}{k} (w+\s) - \frac{1-\hz}{1+\hz} [\l(\t)-\l(\t)^{-1}] \\
\qquad \qquad + 2\log{\big( \frac{\hz\l(\t)-1}{\hz-\l(\t)} \big)} = 0\end{matrix*}$
\vspace{0.5cm}
\end{tabular}

\begin{tabular}{l || l | l | l}
& Symmetric space $\s$-model & Symmetric space $\l$-model \\ \hline\hline

Lagrangian& $\L = -\ha h \Tr [ J_+P_{G/H} J_- ] $ & $\begin{matrix*}[l]\L =k\big[ \L_{G/G}(g,A)\\
\qquad \qquad - \Tr \big[ A_+ (\l^{-1} P_{G/H} -1) A_-\big] \end{matrix*}$\\ \hline

Lax connection &
$\begin{matrix*}[l] L_\pm = J^{H}_\pm + z^{\pm 1} J^{G/H}_\pm \\ J^{H}_\pm = P_H J_\pm , \ J^{G/H}_\pm = P_{G/H} J_\pm \end{matrix*}$
& $\begin{matrix*}[l] L_\pm = A^{H}_\pm + z^{\pm 1} \frac{1}{\sqrt{\l}} A^{G/H}_\pm \\ A^{H}_\pm = P_H A_\pm , \ A^{G/H}_\pm = P_{G/H} A_\pm \\
A_\pm \text{ takes on-shell value} \end{matrix*}$ \\ \hline

\makecell[l]{1-loop RG equation \\
and its solution} & $\begin{matrix*}[l] \ddt h = 2\cg \\
h(t) = 2\cg t
\end{matrix*}$
& $\begin{matrix*}[l]
\ddt\l = \frac{\cg}{k} \l \\
\l(t) = \exp{( \frac{\cg}{k} t )}
\end{matrix*}$
\\ \hline

\makecell[l]{Spectral parameter \\
function}
& $\hz=
\sqrt{\frac{w+\s-\t}{w+\s+\t}}$ &
$ \hz =
\exp{( \frac{\cg}{2k} \t)} \sqrt{\frac{1+\exp{(\frac{\cg}{k} [w+\s-\t] )}}{1+\exp{( \frac{\cg}{k} [w+\s+\t] )}}}
$
\vspace{0.5cm}
\end{tabular}

\caption{\small Examples of integrable \sms and their time-dependent generalizations.
$t$ is the RG time which is replaced by the 2d time $\t$ in the couplings entering the
Lagrangian $\wcL$ in \rf{lcg}.
\label{tab1}}
}
\end{table}

In this section we shall explicitly consider only the basic PCM example,
postponing a more general discussion to section \ref{s3},
where we shall also explain that the existence of the spectral function $\hhz(w;\t,\s)$ is a highly non-trivial feature, depending on
specifically choosing the coupling functions $h_\a(\t)$ to solve the 1-loop RG equation.

For the PCM corresponding to a simple Lie group $G$
\be
\L =-\ha h \Tr[J_+ J_-] \ , \qquad J=g^{-1} dg \ , \ \ g\in G \ , \la{pl}
\ee
the global $G\times G$ symmetry implies
that only the overall coupling $h$ runs under the 1-loop
RG,\foot{\la{fo8}The dual Coxeter number $\cgg$ is defined by
${f^{mn}}_k f_{mnl} = 2\cgg \d_{kl}$, where
$[T_m, T_n] = i {f_{mn}}^k T_k$
and the generators are normalized as $\Tr(T_m T_n) = \d_{mn}$.}
\be
\ddt h = \cg \ ,\qquad \qquad h(t) =\cg t \ , \qquad \qquad \cg \equiv \cgg.
\ee
The corresponding time-dependent theory
\rf{lcg} is then
given
by
\be
\widehat{\L} = - \tfrac{1}{2} \cg \t \, \Tr[J_+ J_-] \ . \la{lcPCM}
\ee
As explained in the Introduction, this theory arises naturally in the l.c.\ gauge $u=\t$ from the conformal string $\s$-model (cf.\ \rf{csm},\rf{lct})
\be
\L =-2 \del_+ u \del_- v - \tfrac{1}{2} \cg u \Tr[J_+ J_-] \ , \la{nPCM}
\ee
where the constant $\cg$ can be eliminated by a rescaling $(u,v) \to ( \cg^{-1} u,\, \cg v)$.

The equations of motion for the model \rf{lcPCM} written in first-order form are
\begin{align}
&\del_+ (\t J_- ) + \del_- (\t J_+) = 0 \ , \la{pe1} \\
&F_{+-}(J) \equiv \del_+ J_- - \del_- J_+ + [J_+, J_-] = 0 \ . \la{pe2}
\end{align}
Starting from the Lax connection of the original PCM \rf{pl}
\be
L_\pm = \ha(1+z^{\pm 1}) J_\pm \ , \la{pla}
\ee
and making the replacement
$z \to \sqrt{\frac{{w+\s-\t}}{{w+\s+\t}}} $ where $w$ is a constant spectral parameter
(see Table \ref{tab1}), we obtain the following expression for the
Lax connection \rf{lac2} of the time-dependent theory
\be
\widehat{L}_\pm = \ha \big( 1+ {{z}\, }^{\pm 1}
\big) J_\pm \ , \qquad \qquad {z}=\sqrt{\frac{{w+\s-\t}}{{w+\s+\t}}} \ . \la{lpl}
\ee
Indeed,
its curvature is
\begin{align}
&F_{+-}(\widehat{L}) \equiv \del_+ \wL_- - \del_- \wL_+ + [\wL_+, \wL_-] \la{curv} \\
&\qquad \ \ \ \, \, = \frac{{z}}{2(w+\s-\t)} \Big[ \del_+ (\t J_- ) + \del_- (\t J_+) \Big]
+ \tfrac{1}{4} (1+{z})(1+{{z}\, }^{-1} ) \ F_{+-}(J)
\ , \no
\end{align}
so that, at
any $(\t, \s)$, its vanishing for all $w$ implies
the equations of motion \rf{pe1},\rf{pe2} of the time-dependent theory.

A surprising feature
(shared also by the other examples in Table \ref{tab1})
is that, despite the Lagrangian \rf{lcPCM} explicitly involving only $\t$,
the Lax connection \rf{lpl}
also depends on the 2d spatial coordinate $\s$.
The spatial coordinate $\s$ and the spectral parameter $w$ appear only through
the combination $w+\s$. Since $w$ is a complex constant, it cannot be
eliminated by a shift of real $\s$. Moreover, what is important is the
existence of a Lax connection depending on $w$, which can in principle be used
to construct conserved charges, etc. Indeed, we can instead interpret the
formal possibility to shift $\s$ as the freedom to introduce a spectral
parameter in the first place.

Another unusual property (shared by all the examples) is that the Lax connection has branch cuts in the spectral $w$-plane
(e.g., for the PCM the branch cuts end at $w=-\s\pm \t$).
Normally, one could simply remove square roots by redefining the spectral parameter (or more formally moving to an appropriate Riemann surface on which the Lax connection is meromorphic).
However, this is not possible since the positions of the branch cuts depend on $(\t,\s)$ and one cannot redefine the spectral parameter in a way depending on $(\t,\s)$ without changing the equations of motion encoded in the zero curvature condition.

In each case in Table \ref{tab1}, one can freely choose any branch of the function $z(w;\t,\s)$ (in the $w$-plane) while still encoding the correct equations of motion in the zero-curvature equation. For example, in the PCM case, one may choose either sign of the square root ${z} =\pm \sqrt{\tfrac{{w+\s-\t}}{{w+\s+\t}}}$ in the Lax connection \rf{lpl}.
All similar square roots in the spectral functions below
have the same branch ambiguity, corresponding to the option to reverse their sign.

\medskip

Applying a general conformal transformation to the action corresponding to \rf{lcPCM},\rf{lpl} we get (cf.\ \rf{nlc})
\begin{align}
& \widehat{\L} =- \tfrac{1}{2}\cg \, \big(f^+( \xi^+) + f^-(\xi^-)\big) \, \Tr[J_+ J_-] \ , \qquad \la{lac} \\
&\widehat{L}_\pm= \ha \big( \, 1+ z^{\pm 1} \big) J_\pm \ , \qquad \quad
z= \sqrt{ \frac{{w- 2f^-(\xi^-)}}{{w+ 2 f^+ (\xi^+)}}} \ . \la{laca}
\end{align}
This theory may equivalently be obtained from the string $\s$-model \rf{nPCM}
by picking up a generalized l.c.\ gauge $u =\t'= f^+(\xi^+) + f^-(\xi^-)$.\foot{
In the special case of $u=a+b \t$,
the limit $a\to 1$, $b\to 0 $ eliminates the explicit $\tau$-dependence and gives back the original PCM \rf{pl}, with \rf{lac} becoming
the original Lax connection in \rf{pla} with a redefined spectral parameter $z\to \sqrt{\frac{{w-2 a_-}}{{w+2 a_+}}}$, where $a_\pm$ are constants satisfying $a_+ + a_- = 1$.}
We note that the theory \rf{lac} is a special case of \rf{ff} where $\GG_{ij}$ is the group-space metric. Indeed, an equivalent expression for the Lax connection
of this `local' PCM (or symmetric space $\s$-model, see Table \ref{tab1})
was originally found in \cite{Breitenlohner:1986um}
where the dependence of the analog of the
spectral parameter on the functions $f^+(\xi^+) $ and $f^-(\xi^-)$ was
discovered.\foot{The spectral function depends separately on $u$ and its dual field $\td u =f^+( \xi^+) - f^-(\xi^-) $, \
$d u= *d\td u $,
and thus separately on $f^+$ and $f^-$ \cite{Breitenlohner:1986um,Nicolai:1991tt}.
In the derivation of the Lax pair for general $u,\td u= f^+( \xi^+) \pm f^-(\xi^-) $ it is clear that $w$ appears as a constant of integration.}

\section{RG flow from condition of integrability of time-dependent theory}\label{s3}

As we have found above on several examples,
if the couplings of an integrable $\s$-model are promoted to functions of time that
solve the 1-loop RG equations, $h_\a \to h_\a(\t),\ \del_\t h_\a = \b_\a(h)$,
then the Lax connection of the original model $\L(h_\a)$ admits a natural generalization
to a classical Lax connection for the time-dependent model $\L(h_\a(\t))$.
Here we shall argue that the converse is also true:
demanding the existence of a Lax pair for the time-dependent theory implies that $h_\a (\t)$ must solve
the 1-loop RG equations.

It is useful to start with a more general theory with local couplings\foot{We generally expect integrable models to have a finite number of couplings $h_\a$.
A natural way to identify these couplings is
as the parameters that run under the RG flow.}
$h_\a(\t,\s)$ depending on both $\t$ and $\s$, i.e.
\begin{align}
&\widehat{\L} = \L(h_\a(\t,\s)) \ , \la{sth}
\end{align}
and demand the existence of a Lax representation for this
theory.
For the examples in Table \ref{tab1}, the original Lax connection
takes a particular form
(which is generic to many integrable $\s$-models built on groups $G$ or symmetric spaces $G/H$):
\begin{alignat}{2}
&\hspace{-2.5cm}G \ : \qquad\qquad & &L_\pm = \ha (1+z^{\pm 1}) \mc{A}_\pm \la{gl} \ ,\\
&\hspace{-2.5cm} G/H \ :
\qquad\qquad & &L_\pm = \mc{B}_\pm + z^{\pm 1} \mc{P}_\pm \la{sl} \ .
\end{alignat}
Here the connection components $\mc{A}_\pm \in \Lie(G)$,
$\mc{B}_\pm \in \Lie(H)$, $\mc{P}_\pm \in \Lie(G)/\Lie(H)$ depend implicitly on the fields and their derivatives
and the couplings $h_\a$ (e.g.\ for PCM, $\mc{A}_\pm = J_\pm $, cf.\ \rf{pla}), while the dependence on the
spectral parameter $z$ is indicated explicitly.
Let us now take the following natural ansatz for the Lax connection $\widehat{L}_\pm$
corresponding to the $(\t,\s)$-dependent theory \rf{sth}
\begin{alignat}{2}
&G \ : \qquad \qquad & &\widehat{L}_\pm = p_\pm(w;\t,\s)\, \mc{A}_\pm\ , \la{glt}\\
& G/H \ : \qquad \qquad & &\widehat{L}_\pm = q_\pm(w;\t,\s)\, \mc{B}_\pm + r_\pm(w;\t,\s)\, \mc{P}_\pm \la{slt} \ ,
\end{alignat}
where $p_\pm, q_\pm, r_\pm$ are some functions, and $\mc{A}_\pm$, $\mc{B}_\pm$, $\mc{P}_\pm$ are assumed to take the same form as in \rf{gl},\rf{sl}, but with the couplings they depend on now being
the functions $h_\a(\t,\s)$, i.e.
$\mc{A}_\pm = \mc{A}_\pm (x,\del x; h_\a(\t,\s))$, etc.

Demanding that the flatness of the connection \rf{glt},\rf{slt}
gives the equations of motion of the generalized theory \rf{sth},
we first prove that it is sufficient to use just
a single function
$\hz(w;\t,\s)$ so that \rf{glt},\rf{slt} become the same as \rf{gl},\rf{sl}
with $z\to \hz(w;\t,\s)$
as in \rf{lac2},
\be
\widehat{L}_\pm = L_\pm\big( h_\a(\t,\s), \hz(w;\t,\s)\big) \ . \la{laxs}
\ee
Indeed, the equations of motion for the theory \rf{sth} contain new
terms proportional to the space-time derivatives of the couplings $h_\a$ as well as
the original terms not depending on $\del_\pm h_\a$.
To get the latter terms from the flatness of $\widehat{L}_\pm$ one needs
to impose the following conditions on $p_\pm$, $q_\pm$, $r_\pm$
(see Appendix \ref{det} for details)
\begin{alignat}{2}
& G \ : \qquad \qquad & &p_+^{-1} + p_-^{-1} = 2 \ , \la{alg}\\
& G/H \ : \qquad \qquad & &q_\pm = 1 \ , \qquad r_+ r_- = 1 \ . \la{als}
\end{alignat}
Setting $\hz(w;\t,\s) = 2 p_+ - 1$ (for models on $G$) and
$\hz(w;\t,\s) = 2 r_+ $ (for models on $G/H$) we conclude that
the Lax ansatze \rf{glt},\rf{slt} become the same as the original expressions \rf{gl},\rf{sl} with $z\to \hz(w;\t,\s)$.

All of our examples are single-coupling theories\foot{We do not promote the WZ level $k$ in PCM$_k$ or $\l$-models to
a function of $(\t,\s)$ since the resulting model
would not be well-defined: starting with the
3d representation of the WZ term and replacing $k H \to k(\t,\s) H$
would no longer give a local 2d action
while $k B \to k(\t,\s) B $ in the 2d action would not be consistent with global symmetry
(invariance of the 2d action under gauge transformations of $B$ requires dropping total derivatives).}
except for the group $\eta$-model. Here we shall
specialize to the simplest case with one coupling $h(\tau,\sigma)$
but the same result will also be true for the group $\eta$-model (see Appendix \ref{mr}).
To reproduce the derivative $\del_\pm h$ terms in the
equations of motion for \rf{sth} from the flatness condition for the generalized Lax connection \rf{laxs}, one needs to additionally impose certain constraints on both $h(\t,\s)$ and $\hz(w;\t,\s)$.
For the models in Table \ref{tab1}
this leads to the following two first-order differential equations for the function $\hz(w;\t,\s)$
(here $h=h(\t,\s)$)
\begin{align}
\del_\t z = V_\t (z,h) \ , \qquad\qquad \del_\s z = V_\s (z,h) \ , \la{pd1}
\end{align}
where $V_{\t,\s} (z,h)$ are model-dependent functions. For all five single-coupling examples in Table \ref{tab1}, the consistency condition ($\del_\t V_\s - \del_\s V_\t =0$) for the system \rf{pd1} takes the
remarkable form
\be
\del_+ \del_- h - \frac{\b'(h)}{\b(h)}\, \del_+ h \del_- h = 0 \la{2w} \ ,
\ee
where $\b(h)$ is precisely the 1-loop $\b$-function for the coupling $h$ in the original model.
In addition
to the appearance (in this purely classical context)
of the 1-loop $\beta$-functions, a remarkable feature of \rf{2w}
that it does not depend on $z$ (which completely factors out of the consistency condition).

The condition \rf{2w} may be written simply as $ \del_- ( \frac{\del_+ h}{\b(h)}) =0$ or
$ \del_+ ( \frac{\del_- h}{\b(h)}) =0$ and thus leads to two first-order equations
\be
\frac{\del_+ h}{\b(h)} = s_+(\xi^+) \ , \qquad \qquad \frac{\del_- h}{\b(h)} = s_-(\xi^-) 	\ , \la{cfs}
\ee
where $s_\pm (\xi^\pm)$ are arbitrary functions of $\xi^\pm =\ha ( \t \pm \s)$.
By applying a conformal transformation (i.e.\ redefining $\t$ and $\s$), one can absorb $s_\pm$ into $\del_\pm$, i.e.\ replace
$s_\pm \to 1$,
so that in terms of the redefined coordinates the first-order equations \rf{cfs}
take the form of the 1-loop RG equation in $\tau$
\be
\del_\t h = \b(h) \ , \qquad \qquad \del_\s h = 0 \ . \la{RGs2}
\ee
Thus, modulo a conformal transformation,
the 1-loop RG solution is the only choice
of local coupling
$h(\t,\s)$
for which the Lax connection can be uplifted to the $(\t,\s)$-dependent theory according to \rf{laxs}.
This argument (in eqs. \rf{pd1}-\rf{RGs2}) is demonstrated explicitly for the example of the PCM in Appendix \ref{fdr}.

The same conclusion is reached of course if one starts directly with the theory where
the couplings $h_\a$ depend only on time:
using the same ansatz for the Lax connection
implies that the only functions $h_\a(\t)$ that are consistent with preserving integrability
are the solutions of the 1-loop RG flow (in this case in
eqs.\ \rf{2w},\rf{cfs} we have $\del_\pm \to \del_\t$
and $s^+=s^-=\const$).\foot{This is true modulo
a rescaling of time, i.e.\ the part of the conformal group that does not introduce spatial dependence.
Note that the freedom of performing a classical conformal transformation means that instead of assigning a preferred role to $\t$ we could have chosen the couplings to depend only on $\s$; in this case the RG equation \rf{RGs2} will hold with $\t \to \s$.}

To finish the construction of the generalized Lax pair let us now explain how to obtain the explicit form of the spectral parameter function
$z(w;\t,\s)$ in \rf{laxs}.
Starting with $h=h (\t)$ that satisfies the 1-loop RG equation in \rf{RGs2}
(so that the consistency conditions \rf{cfs} of \rf{pd1}
are satisfied) one can solve the second equation in \rf{pd1} as
\be
\del_\s z = V_\s (z, h(\t)) \qquad \rightarrow \qquad \int \frac{dz}{V_\s(z,h_\a(\t))} =\s + \ell (\t) \ . \la{sold}
\ee
The function $\ell(\t)$ is then fixed by substituting this solution into the first equation in \rf{pd1}.
The solution for $\ell(\t)$ leaves one free integration constant, which we call
$w$. Finally, eq.~\rf{sold} becomes an algebraic equation for $z(w;\t,\s)$
that can be explicitly solved in some simple cases (see Table \ref{tab1}).
Note that since the parameter $w$ appears as an integration constant in the function $\ell(\t)$,
it always appears in the combination $w+\s$ with the spatial coordinate,
implying that constant shifts of $\s$ can be compensated by shifts of $w$.

Although we do not have a proof that the 1-loop RG flow always follows from requiring
integrability of the time-dependent generalizations of integrable models, the examples
in Table \ref{tab1} reveal a highly non-trivial pattern suggesting that this may be true for a more
general class of theories (see also section \ref{s6}).\foot{Notice, in particular, that
some of the 1-loop RG equations in Table \ref{tab1} are quite non-trivial, not admitting simple closed form solutions.}

\section{Conserved charges in time-dependent integrable models}\la{s4}

We have shown in the previous sections that certain well-known examples of classically integrable $\s$-models admit generalizations of their Lax connections to models with time-dependent couplings under the condition that the latter solve
the 1-loop RG equation in $\tau$.
However, one may question the usefulness of the resulting Lax connections since they have several unusual properties: they depend on $\t$ and $\s$ explicitly; the spectral parameter only enters as a `constant piece' of the spatial coordinate $\s$; they
have branch cuts in the spectral plane (whose positions depend on $\t$ and $\s$).

In this section we shall argue that, nevertheless, the Lax connections can still be used to construct the `non-local' charges typical of integrable models. However, the space-time dependence of the Lax connection renders these charges difficult to compute explicitly. We shall focus in this section on the case of the PCM, while the other examples from Table \ref{tab1} are discussed in Appendix \ref{om}.

\subsection{Non-local charges}\la{s41}

Let us start by reviewing the standard construction of non-local charges using a Lax connection. On a given spatial domain $a<\s<b$ the \textit{monodromy matrix} is defined by
\begin{align}
&\M(\t) \equiv P\exp \int_a^b d\s \, L_\s(\t,\s) \ , \la{p1} \\
&\M^{-1}\del_b \M = L_\s(b) \ , \qquad \M\big|_{b=a} = I \ , \qquad \qquad
L_{\t,\s}(b) \equiv L_{\t,\s}(\t,\s)\big|_{\s=b} \ . \la{px}
\end{align}
Using the flatness of the Lax connection, $\del_\s L_\t - \del_\t L_\s + [L_\s,L_\t]=0$,
it easy to check that
\be
\del_\t \M = \M \, L_\t(b) - L_\t(a) \, \M \ . \la{ce}
\ee
Thus, if we assume the periodicity condition
\be
L_\t(b) = L_\t(a)\ , \la{pe}
\ee
we obtain
\begin{align}
\del_\t \M = [\M, L_\t(a)] \ . \la{lp}
\end{align}
Hence it follows that $\del_\t \Tr [\M^n]=0$ for all $n$, or, equivalently, the eigenvalues of $\M$ are conserved in time.

We shall focus on the theory on an infinite spatial interval $(a,b)=( -\infty,\infty)$.\foot{It seems much harder to satisfy the periodicity condition \rf{pe} on a spatial circle due to the explicit non-periodic $\s$ dependence in the Lax connections for the time-dependent models.}
Picking the `negative' branch\foot{We note that it is consistent to `pick a branch' here: e.g., by assuming
that $w$ has an imaginary part, it then follows that the sign of the square root does not change from $\s=-\infty$ to $\s=\infty$. The choice of a branch is arbitrary but this negative choice is more useful for satisfying the periodicity condition and constructing conserved charges. Also, for the group $\eta$-model and PCM$_k$, which are deformations of PCM, the function
${z}$ is single-valued at $\s=\pm \infty$, with $z_{\infty}\equiv z\big|_{\s=\infty}=-1$. It is then natural to obtain $z_{\infty}=-1$
for the PCM as a limit of these models.} of the square root,
then the spectral function ${z}(w;\t,\s)$ in \rf{lpl} has the finite limit ${z} =\sqrt{\frac{w+\s+\t}{w+\s-\t}} \to z_{\infty} = -1$ at $\s \to \pm \infty$.
Hence, assuming $J_\pm$ is bounded for large $|\s|$, the Lax connection $\widehat{L}_\pm = \ha (1+{z}^{\pm 1} ) J_\pm$
vanishes at spatial infinity. The periodicity condition \rf{pe} is then satisfied so it follows from \rf{lp} that the eigenvalues of $\M$ are conserved.
Furthermore, since $\widehat{L}_\t$ actually \textit{vanishes} at spatial infinity,
it follows from \rf{ce} that all the components of $\M$ are conserved.

The boundary condition (that $J_\pm$ is bounded), needed above for the conservation of $\M$, is quite weak. A further, more stringent boundary condition comes from the requirement that $\M$ converges
as $(a,b)\to (-\infty, +\infty)$
to produce well-defined charges.

Although the monodromy is conserved, it is only defined implicitly by the first-order ordinary differential equation \rf{px}, i.e.
\begin{align}
&\M^{-1} \del_\s \M = \widehat{L}_\s = \tfrac{1}{4}\big(1+\big[ \tfrac{w+\s-\t}{w+\s+\t} \big]^{1/2} \big)J_+ - \tfrac{1}{4}\big(1+\big[ \tfrac{w+\s-\t}{w+\s+\t} \big]^{-1/2}\big)J_- \ , \qquad \M\big|_{\s=-\infty} = I \ . \la{mpe}
\end{align}
Such an equation generally admits a solution but it is hard to find its explicit form. One possible approach
is to develop an expansion in the spectral parameter around a point where $\wL_\s$ vanishes (and thus the monodromy is trivial, $\M=I$).

In the usual time-independent PCM case (obtained by replacing $\sqrt{\tfrac{w+\s-\t}{w+\s+\t}} $ by a constant $z$ in \rf{mpe}), such an expansion around $z=-1$ yields the conserved `multi-local' Yangian charges,
\begin{align}
&\M =I + \ha (z+1) \int_a^b d\s \, J_\t \la{me} \\
&\qquad\ \ \ \
+ \tfrac{1}{4} (z+1)^2 \Big[ \int_{a<\s_1<\s_2<b} d\s_1 d\s_2 \, J_\t(\s_1)J_\t(\s_2) + \int_a^b d\s \,(J_\t-J_\s) \Big]
+ \O\big((z+1)^3\big) \no
\ .
\end{align}
Each term in the expansion \rf{me} is individually conserved because each term in the corresponding expansion of $L_\t$ (and hence the right hand side of \rf{ce}) vanishes at spatial infinity.

For the time-dependent theory, the only zero around which to expand $\widehat{L}_\s$ in \rf{mpe} is $w=\infty$ (again taking the negative branch of the square root). But, due to its explicit spatial dependence, the corresponding expansion of the Lax connection
is a sum of terms $w^{-n} P_{n-1}$, where $P_{n-1}$ is a polynomial of degree $n-1$ in $\s$. Then for any (polynomial) decaying boundary conditions on the fields at spatial infinity, the periodicity condition \rf{pe} on $\widehat{L}_\t$ will be broken at sufficiently higher order in this expansion.

Hence, while we have shown the formal existence of conserved non-local charges defined implicitly by \rf{px}, it appears to be hard to compute the monodromy matrix $\M$ explicitly: due to the explicit $\s$ dependence in the Lax connection, the usual expansion trick \rf{me} does not work.
It is not clear at the moment how
to verify
if the resulting conserved charges are infinite in number (and independent)
as they should be for an integrable 2d theory.

\subsection{1d reductions}\la{s43}

One useful test of 2d integrability is to check the integrability of various 1d mechanical theories obtained as
consistent reductions of the 2d equations of motion.

Assuming an ansatz for a classical solution of an integrable 2d model
yielding a 1d system of equations (say with $\t$ as the remaining variable), one may expect
the resulting 1d system also to be integrable in the sense of admitting a Lax pair, $\tfrac{d}{d\t} A = [B, A]$. One should further demand that the conserved charges are in involution.

In general, if the solution satisfies the periodicity condition \rf{pe}, the flatness of the 2d Lax connection leads to a 1d Lax pair
given by $(A,B)=(\M, -L_\t(\infty))$ in \rf{lp}, evaluated on the reduction ansatz.\foot{In \cite{Arutyunov:2003za} such a reduction was performed for a certain `spinning string' ansatz in $AdS_5 \times S^5$, resulting in an integrable 1d Neumann-Rosochatius model. There (reversing the roles of $\t$ and $\s$ from that paper to relate to the above discussion) the $\s$ dependence of the ansatz was simple and it was possible to remove the $\s$ dependence from the Lax connection using a gauge transformation -- which is essentially the same problem as computing the monodromy matrix. In our case, with explicit $\s$-dependence already in the Lax connection, the problem of removing the $\s$ dependence using gauge transformations appears to be much more non-trivial.}

\paragraph{Trivial reduction.}
In the simplest `trivial' 1d reduction one assumes that the
fields do not depend on $\s$, which is a consistent truncation of the above time-dependent models.
For example, in the PCM case setting $g=g(\tau)$ in \rf{lcPCM} gives the 1d action for
geodesic motion on a group space with a time-dependent radius.
The solvability of geodesics in this model is
obvious since the explicit time dependence may be eliminated by a redefinition of $\t$
\be
\la{48} S_1 \sim \int d\t \ \t \ \Tr\big[ (g^{-1} \del_\t g )^2 \big] = \int d\t' \ \Tr\big[ (g^{-1} \del_{\t'} g )^2 \big] \ , \qquad \ \ \t'=\log \t \ .
\ee
The resulting equation of motion $\del_{\t'} (g^{-1} \del_{\t'} g) = 0$ is solved by
\be
g = g_0\, e^{\t' u_0} = g_0\, \t^{u_0 } \ , \qquad \quad g_0 = \const \in G \ , \quad \ u_0=\const \in \Lie(G) \ .
\ee
The same argument equally applies for any case where the time dependence only appears as an overall factor rescaling the Lagrangian -- which will follow if there are enough global symmetries that only the overall `radius' can run under the RG flow (e.g.\ also in the symmetric space $\s$-model case).

In general, the global symmetry charges may be found from the monodromy $\M$. Formally, the monodromy on an infinite line does not converge upon the trivial reduction since $J_\t = g^{-1} \del_\t g$ does not decay at spatial infinity (as it is now independent of $\s$).
On a finite space interval, the periodicity condition \rf{pe} is not satisfied so charges will not be conserved. However, making an expansion around the zero of the Lax connection at $w=\infty$ (cf.\ \rf{me}), the global symmetry is reflected in the $\s$ independence of the leading $\O(w^{-1})$ term in $\wL_\t$,
\begin{align}
\wL_\t &= \tfrac{1}{4}(1+\big[ \tfrac{w+\s-\t}{w+\s+\t} \big]^{1/2})J_+ + \tfrac{1}{4}(1+\big[ \tfrac{w+\s-\t}{w+\s+\t} \big]^{-1/2})J_-
= \tfrac{1}{2} \t J_\s \, w^{-1} + \O(w^{-2})\ .
\end{align}
This puts the leading term on the right hand side of \rf{ce} in the commutator form \rf{lp}. Since the monodromy matrix is the identity at the leading order,
the right hand side of \rf{ce} is of $\O(w^{-2})$. It then follows that the $\O(w^{-1})$ term in the monodromy matrix, which is the global $G_R$ Noether charge, is
still conserved\foot{The Noether charge for $G_L$ symmetry is obtained similarly after a gauge transformation of the Lax connection and choosing instead the positive branch of the square root.}
\begin{align}
&\M = 1 + \ha w^{-1} Q_R + \O(w^{-2}) \ , \qquad \del_\t Q_R = 0 \ , \qquad Q_R = \int d\s \ \t J_\t \ . \la{noet}
\end{align}

\paragraph{Non-trivial reduction.}
Now let us consider a non-trivial 1d reduction when the 2d fields \textit{do} depend on $\s$, but in a particular prescribed way. Such a reduction will lead to a
non-trivial 1d model, giving a more stringent test of the integrability of the time-dependent 2d model.

Starting from the time-dependent $SU(2)$ PCM \rf{lcPCM} parametrized as
\begin{align}
&\wLL = -\tfrac{c}{2} \, \t \, \Tr [ J_+ J_- ] = c\, \t \, \big[ \del_+ \theta \del_- \theta + \sin^2{\theta} (\del_+ \phi \del_- \phi + \sin^2{\phi}\, \del_+ \psi \del_- \psi)\big] \ , \la{412} \\
&\qquad g = n^a \s_a \ , \qquad \s_a = (I, \s_i ) \ , \ \ i=1,2,3 \ , \no \\
&\qquad n^0 n^0 - n^i n^i = 1 \ , \qquad n^a = ( \cos{\theta} , \ i \sin{\theta} \cos{\phi} , \ i \sin{\theta} \sin{\phi} \cos{\psi} , \ i \sin{\theta} \sin{\phi} \sin{\psi} ) \ , \no
\end{align}
let us consider, e.g., the following ansatz for a classical solution
\be
\theta = \theta(\t) \ , \qquad \phi = m \s \ , \qquad \psi = 0 \ . \la{wan}
\ee
This leads to a consistent reduction of the 2d theory: the $\phi$ and $\psi$ equations are both satisfied,
while the equation for $\theta(\t) $ follows from the 1d Lagrangian ($\dot{\theta} \equiv \del_\t \theta$)
\be
\L_1 = {c} \, \t\, \big( \dot{\theta}^2 - m^2 \sin^2 {\theta}\big) \ . \la{s1}
\ee
Thus the effective 1d model is a time-dependent analog of the `sine-Gordon' mechanics.

Let us assume the $\s$-direction is an infinite line, still using the ansatz \rf{wan} (e.g.\ by taking $m$ to be a continuous parameter and formally decompactifying $\phi$).
Then
one can check that $J_\pm$ is oscillating but bounded at spatial infinity. Hence, as discussed above, the right hand side of \rf{ce} vanishes and the entries of the monodromy matrix are conserved.
While again it is not easy to compute the monodromy explicitly in terms of $\theta$ and $\dot{\theta}$,
it can be done in the small field expansion $(\theta,\dot{\theta}) \to \e (\theta,\dot{\theta})$, $\e \ll 1$ since the Lax component $\wL_\s$ vanishes as $(\theta, \dot{\theta}) \to 0$. In this expansion the Lagrangian \rf{s1} becomes (after rescaling it by a factor of $c^{-1} \e^{-2}$)
\begin{align}
&\L_1' = \t\big[ \dot{\theta}^2 - m^2 \e^{-2} \sin^2 {(\e \theta)}\big] = \L^{lin} + \tfrac{1}{3} \e^2\, m^2\, \theta^4 + \O(\e^4) \ , \la{ne}\\
&\L^{lin} = \t \, \big[ \dot{\theta}^2 - m^2 \theta^2 \big] \ . \la{lt}
\end{align}
The leading (`linearized') Lagrangian \rf{lt} describes (after the same redefinition $\tau = e^{\tau'}$ as in \rf{48})
a harmonic oscillator with time-dependent frequency $m^2 e^{2 \tau'}$.
The corresponding $\O(\e)$ terms in the monodromy matrix lead to the conserved charge
\begin{align}
&\dot{Q}^{lin} = 0 \ , \qquad Q^{lin} = \t \big[ \bw(\t; w) \, \dot{\theta} - \dot \bw(\t; w) \, \theta \big] \ , \la{ql} \\
& \bw(\t;w) = \int_{-\infty}^{+\infty} d\s\, e^{-im \s} \tfrac{ i}{2 s_+ s_- }
\ , \qquad \qquad s_\pm \equiv \sqrt{w+\s \pm \t}\ . \la{ql2}
\end{align}
In Appendix \ref{C} we use this conserved charge to construct the general solution of the linearized equation of motion
(with $\theta=\bw$ being a particular solution).

Computing the perturbative corrections to the monodromy, one then finds for the
conserved charge of the non-linear theory \rf{ne}
\begin{align}
&\dot{Q} = 0 \ , \ \ \ Q = Q^{lin} + \e^2 \Big[\int d\s \, q^{(1)} + \int_{\s_1<\s_2} d\s_1 d\s_2 \, q^{(2)} + \int_{\s_1<\s_2<\s_3} d\s_1 d\s_2 d\s_3 \, q^{(3)} \Big] + \O(\e^4) \ , \no \\
&\ \ \ q^{(1)} = - \tfrac{m \, e^{-im\s} }{3s_+ s_-} (s_+ s_- -w-\s) \, \theta^3 \ , \la{cq}\\
&\ \ \ q^{(2)} = \tfrac{i m }{4s^1_+ s^1_-s^2_+ s^2_-} \Big[ e^{-im \s_2} ( s_+^1 s_-^1 - w - \s_1)\big[m(s_+^2 s_-^2 -w- \s_2)\theta +i \t \dot{\theta}\big] - (\s_1 \leftrightarrow \s_2) \Big] \, \theta^2\ , \no \\
&\ \ \ q^{(3)} = \tfrac{e^{-im(\s_1 - \s_2 + \s_3)}}{8 s^1_+ s^1_-s^2_+ s^2_-s^3_+ s^3_-} \big[m(s_+^1 s_-^1 -w- \s_1)+i \t \dot{\theta}\big] \no \\
&\qquad \qquad \times \big[ - m(s_+^2 s_-^2 -w- \s_2)+i \t \dot{\theta}\big]\big[m(s_+^3 s_-^3 -w- \s_3)+i \t \dot{\theta}\big] \ , \ \ \quad s_\pm^n \equiv \sqrt{w+\s_n \pm \t} \ . \no
\end{align}
This perturbative procedure of constructing the conserved charge can be extended to higher orders,
computing more and more $\s$ integrals at each step.

Let us mention that the time-dependent $SL(2,\mathbb{R})$ PCM (cf.\ \rf{412})
\be\la{420}
\wLL = - c \, \t \Big[ \del_+ \r \del_- \r + \ha e^{-2\r} \, (\del_+ x^+\del_- x^- + \del_- x^+\del_+ x^-) \Big] \ ,
\ee
admits a `non-trivial' reduction that is manifestly integrable.
Indeed, if we set $x^+ = a \s, \ x^- = b\s, \ \r=\r(\t)$ then the $x^\pm$ equations are solved
and the equation for $\r(\t)$ follows from the 1d action (cf.\ \rf{s1})
\be\la{421}
S_1 = - c \,\int d \t \, \t \big[ (\del_\t{\r})^2 - ab \, e^{-2\r} \big] \ .
\ee
Here the explicit time dependence can be eliminated by redefining $\t=e^{\t'}$ and $\r = \r' + \t'$
so that we end up with the standard 1d Liouville mechanics with the energy being a conserved charge.

\section{Hamiltonian formulation and local conserved charges}\label{s5}

While the Lax connection is suggestive of integrability, we would like to establish the existence, or otherwise, of an infinite tower of local conserved charges in involution (cf. \cite{Goldschmidt:1980wq,Evans:1999mj,Evans:2000qx,Lacroix:2017isl})
to put the status of the models with local couplings on a firmer footing.
Here we investigate this question using the Hamiltonian formulation for the PCM with local couplings.
While we will only be able to construct such conserved charges for the `chiral' theories,
where the coupling depends on a
l.c.
coordinate $\xi^-$ (see also Appendix \ref{chiral}),
the resulting Lax matrix algebra is suggestive of an underlying algebraic structure that remains to be understood
further.

We start by considering the PCM with only space-dependent coupling
\begin{equation}\label{eq:lagsig}
\widehat{\mathcal{L}} = - \tfrac{1}{2} (1+ b\, \sigma)\, \Tr[J_+J_-] \ ,
\end{equation}
which corresponds to choosing $f^\pm(\xi^\pm) = \frac{1}{2} \pm b\, \xi^\pm$ in the general Lagrangian~\eqref{lac}
(we ignore the overall constant, setting $c=1$). The standard PCM is recovered for $b=0$.
This choice has the advantage that the Hamiltonian governing the time evolution of the model
is a conserved charge.
We will again focus on the theory on an infinite spatial interval $\sigma \in (-\infty,\infty)$ and assume that the boundary conditions decay sufficiently fast to neglect boundary terms.

To develop the Hamiltonian formulation we follow \cite{Faddeev:1985qu} (see also \cite{Delduc:2019bcl} for a more modern treatment).
We parametrize the group-valued field in terms of coordinates $\phi^m$ and write $g^{-1}\partial_m g = i E_m{}^n(\phi_k) T_n$.
Recalling that in our conventions $\Tr[T_mT_n] = \delta_{mn}$, the Lagrangian \eqref{eq:lagsig} in terms of the fields $\phi^m$ is given by
\begin{equation}\label{eq:lagsig1}
\widehat{\mathcal{L}} = \tfrac12 (1+b\,\s) E_m{}^nE_k{}^l \delta_{nl}\, \partial_+ \phi^m \partial_- \phi^k \ ,
\end{equation}
where we use $\delta_{mn}$ and its inverse to lower and raise indices.
The conjugate momenta are
\begin{equation}\label{eq:pi}
\pi_m = \frac{\partial\widehat{\mathcal{L}}}{\partial(\partial_\tau\phi^m)} = (1+b\,\s) E_m{}^n E_k{}^l \delta_{nl}\, \partial_\tau \phi^k \ ,
\end{equation}
and the equal-time Poisson brackets take the standard form
\begin{equation}
\{\phi^m (\sigma_1),\phi^n(\sigma_2)\} = 0 \ , \qquad
\{\pi_m (\sigma_1),\pi_n(\sigma_2)\} = 0 \ , \qquad
\{\phi^m (\sigma_1),\pi_n(\sigma_2)\} = \delta^m{}_n \delta_{\sigma_1\sigma_2} \ ,
\end{equation}
where $\delta_{\sigma_1\sigma_2} = \delta(\sigma_1 - \sigma_2)$.

We define
\begin{equation}\la{jX}
X = i (E^{-1})_n{}^m \pi_m \delta^{nk} T_k \ ,
\end{equation}
such that on \eqref{eq:pi} we have $X = (1+b\,\s) J_\t= (1+b\,\s) g^{-1}\partial_\tau g$.
The Poisson brackets of $g$ and $X$ are
\unskip\footnote{The notation here is the standard one (see, e.g., \cite{Faddeev:1985qu}). To compactly write the Poisson brackets of matrix-valued fields $A$ and $B$ we define $A_\ei{1} = A \otimes 1$ and $B_\ei{2} = 1 \otimes B$. The Poisson bracket $\{A_\ei{1}(\sigma_1),B_\ei{2}(\sigma_2)\}$ is $\{A^{ij}(\sigma_1),B^{kl}(\sigma_2)\}e^{(1)}_{ij}\otimes e^{(2)}_{kl}$, where $e^{(1)}_{ij}$ and $e^{(2)}_{kl}$ are bases for the spaces in which the respective fields are valued.}
\begin{equation}\begin{aligned}
\{g_\ei{1}(\sigma_1),g_\ei{2}(\sigma_2)\} &= 0 \ ,
\qquad
&\{X_\ei{1}(\sigma_1),X_\ei{2}(\sigma_2)\} &= [C_\ei{12},X_\ei{1}(\sigma_1)]\delta_{\sigma_1\sigma_2} \ ,
\\
\{g_\ei{1}(\sigma_1),X_\ei{2}(\sigma_2)\} &= - g_\ei{1}(\sigma_1) C_{\ei{1}\ei{2}} \delta_{\sigma_1\sigma_2} \ ,
\qquad
&\{X_\ei{1}(\sigma_1),g_\ei{2}(\sigma_2)\} &= g_\ei{2}(\sigma_2) C_{\ei{1}\ei{2}} \delta_{\sigma_1\sigma_2} \ ,
\end{aligned}\end{equation}
where $C_{\ei{1}\ei{2}} = \delta^{mn} T_m \otimes T_n$ is the split quadratic Casimir, which obeys
\begin{equation}
[C_{\ei{1}\ei{2}}, A_\ei{1} + A_\ei{2}] = 0 \ .
\end{equation}
It will also be useful to know the Poisson brackets of $j\equiv J_\sigma = g^{-1}\partial_\sigma g$ with $X$
\begin{equation}\begin{aligned}
\{j_\ei{1}(\sigma_1),j_\ei{2}(\sigma_2)\} &= 0\ , \qquad
\{X_\ei{1}(\sigma_1),X_\ei{2}(\sigma_2)\} = [C_{\ei{1}\ei{2}},X_\ei{1}(\sigma_1)]\delta_{\sigma_1\sigma_2} \ ,
\\
\{j_\ei{1}(\sigma_1),X_\ei{2}(\sigma_2)\} &=
\{X_\ei{1}(\sigma_1),j_\ei{2}(\sigma_2)\} = [C_{\ei{1}\ei{2}},j_\ei{1}(\sigma_1)]\delta_{\sigma_1\sigma_2} - C_{\ei{1}\ei{2}} \delta'_{\sigma_1\sigma_2} \ ,
\end{aligned}\end{equation}
where $\delta'_{\sigma_1\sigma_2} = \partial_{\sigma_1}\delta(\sigma_1-\sigma_2)$.

We are interested in the Poisson bracket of the Lax matrix, i.e.\ the spatial component of the Lax connection, with itself.
From the general form of the Lax connection in eq.~\eqref{lac} we find that the Lax matrix, written in terms of $X$ and $j$, is given by
\begin{equation}\la{laj}
\hspace{-0.1cm}\widehat{L}(w;\sigma) \equiv \tfrac{1}{2} (\widehat{L}_+ - \widehat{L}_-) = \frac{ z + z^{-1}+2}{4}\, j + \frac{z - z^{-1}}{4(1+b\,\s)} \, X \ , \ \ \ \ \ \ \
z(w;\sigma) = \sqrt{\frac{w+b\tau -1 - b\,\s}{w+b\tau + 1 + b\,\s}} \ . \end{equation}
Denoting $z_{i} = z(w_{i};\sigma_{i})$ the Lax matrix algebra is then given by
\begin{align}
\{\widehat{L}_\ei{1}(w_1;\sigma_1),\widehat{L}_\ei{2}(w_2;\sigma_2)\}
& = \frac{(z_1 -z_1^{-1})(z_2-z_2^{-1})}{16(1+b\,\s_1)^2}[C_{\ei{1}\ei{2}},X_\ei{1}(\sigma_1)]\delta_{\sigma_1\sigma_2} \no
\\
&\quad + \frac{(1+z_1)(1+z_2)(1-z_1^{-1}z_2^{-1})}{8(1+b\,\s_1)}[C_{\ei{1}\ei{2}},j_\ei{1}(\sigma_1)]\delta_{\sigma_1\sigma_2} \label{laxalg}
\\
&\quad - \Big(\frac{(2+z_1+z_1^{-1})(z_2-z_2^{-1})}{16(1+b\,\s_2)} + \frac{(2+z_2+z_2^{-1})(z_1-z_1^{-1})}{16(1+b\,\s_1)}\Big)C_{\ei{1}\ei{2}}\delta'_{\sigma_1\sigma_2} \ .\no
\end{align}
This bracket is of the type considered in \cite{Maillet:1985ek} (although note that in \cite{Maillet:1985ek} the dependence on $\sigma$ was understood to be through the fields of the model). We can recast eq.~\eqref{laxalg} into the form
\begin{equation}\begin{aligned}\label{laxalgr}
\{\widehat{L}_\ei{1}(w_1;\sigma_1),\widehat{L}_\ei{2}(w_2;\sigma_2)\}
& =
[R_{\ei{1}\ei{2}}(w_1,w_2;\sigma_1),\widehat{L}_\ei{1}(w_1;\sigma_1)]\delta_{\sigma_1\sigma_2}
\\ & \quad
-
[R_{\ei{2}\ei{1}}(w_2,w_1;\sigma_2),\widehat{L}_\ei{2}(w_2;\sigma_2)]\delta_{\sigma_1\sigma_2}
\\ & \quad
- \big(R_{\ei{1}\ei{2}}(w_1,w_2;\sigma_2)+R_{\ei{2}\ei{1}}(w_2,w_1;\sigma_1)\big)\delta'_{\sigma_1,\sigma_2} \ ,
\end{aligned}\end{equation}
with the $R$-matrix given by
\unskip\footnote{To derive this form one can use the identity
\begin{equation*}
F(\sigma_1,\sigma_2)\delta'_{\sigma_1\sigma_2}
= \bigg(\int_0^{\sigma_1} d\sigma_1' \, \Big[ \partial_{\sigma_1'} F(\sigma_1',\sigma_2') \big|_{\sigma_2'=\sigma_1'}\Big]
+ \int_0^{\sigma_2} d\sigma_2' \, \Big[\partial_{\sigma_2'} F(\sigma_1',\sigma_2') \big|_{\sigma_1'=\sigma_2'}\Big]
+ F(0,0)\bigg)\delta'_{\sigma_1\sigma_2} \ ,
\end{equation*}
to represent the coefficient of $\delta'_{\sigma_1\sigma_2}$ in \eqref{laxalg} as a sum of two functions, one depending only $\sigma_1$ and the other only on $\sigma_2$.}
\begin{equation}\label{rmat}
R_{\ei{1}\ei{2}}(w_1,w_2;\sigma) =
\frac{(1+z_1)(1+z_2)(2-z_2 - z_2{}^{-1})}{8(z_1-z_2)(1+b\,\s)}\Big|_{\sigma_i = \sigma} C_{\ei{1}\ei{2}} \ .
\end{equation}
For $b = 0$ the $R$-matrix is non-dynamical (independent of $\sigma$) and we recover the standard result for the PCM \cite{Faddeev:1985qu} in our parametrization.
In this case it is well known that $R_{\ei{1}\ei{2}}$ satisfies the classical Yang-Baxter equation (cYBE)
\begin{equation}\begin{split}\label{eq:cybe}
[R_{\ei{1}\ei{2}}(w_1,w_2;\sigma),R_{\ei{1}\ei{3}}(w_1,w_3;\sigma)]
& +
[R_{\ei{1}\ei{2}}(w_1,w_2;\sigma),R_{\ei{2}\ei{3}}(w_2,w_3;\sigma)]
\\ & \qquad\qquad
+
[R_{\ei{3}\ei{2}}(w_3,w_2;\sigma),R_{\ei{1}\ei{3}}(w_1,w_3;\sigma)] = 0\ .
\end{split}\end{equation}
For $b \neq 0$ the $R$-matrix becomes dynamical, i.e.\ it depends on $\sigma$.
Nevertheless, since all the $R$-matrices in \eqref{eq:cybe} depend on the same $\sigma$, and the $\sigma$ dependence in the $R$-matrix \eqref{rmat} takes a particular form -- it only comes through $z_i$ and an overall factor -- it follows that the cYBE remains satisfied.
This is a sufficient condition for the Lax matrix algebra \eqref{laxalg} to satisfy the Jacobi identity \cite{Maillet:1985ek}.

For the standard PCM with constant coupling one can construct a tower of higher-spin local conserved charges in involution \cite{Evans:1999mj}, which is one of the hallmarks of integrability.
In \cite{Lacroix:2017isl} this construction was generalized to integrable systems with non-dynamical $R$-matrices that can be written in
a `twist' form \cite{Sevostyanov:1995hd}
\begin{equation}\label{eq:tf}
\mathcal{R}_{\ei{1}\ei{2}}(u_1,u_2) = \frac{\varphi(u_2)^{-1}}{u_2-u_1} C_{\ei{1}\ei{2}} \ .
\end{equation}
Indeed, for $b=0$
the $R$-matrix \eqref{rmat} can be written in this form with the spectral parameters and the twist function given by
\begin{equation}
w_i = -\frac{1+u_i^2}{2u_i} \ , \qquad\qquad \varphi(u) = \frac{1-u^2}{u^2} \ .
\end{equation}
Local conserved charges are then given by \cite{Evans:1999mj,Lacroix:2017isl}
\begin{equation}\label{charges}
Q^{(n)}_{\pm1} = \operatorname{res}_{_{u = \pm 1}} \int d\sigma \, \Tr\big[ \varphi(u)^{n-1}\widehat{L}(u;\sigma)^n\big] \ ,
\end{equation}
where $u=\pm1$ are zeroes of the twist function $\varphi(u)$.

For $b\neq 0$, if we allow only redefinitions of the spectral parameter that do not depend on the 2d coordinates,
then it is straightforward to see that it is no longer possible to represent \rf{rmat} in the `twist' form \rf{eq:tf}.
Motivated by standard PCM ($b=0$)
case, we can come somewhat close by introducing
$u(w;\sigma) = \frac{\sqrt{w+b\tau-1-b\,\s}-\sqrt{w+b\tau+1+b\,\s}}{\sqrt{w+b\tau-1-b\,\s}+\sqrt{w+b\tau+1+b\,\s}}$ and $u_i = u(w_i;\sigma_i)$,
in terms of which the Lax matrix and the $R$-matrix are given by
\begin{equation}\begin{split}
\widehat{L}(w;\sigma) & = \frac{1}{1-u^2}j+ \frac{1}{1+b\,\s}\frac{u}{1-u^2} X \ ,
\\
\mathcal{R}_{\ei{1}\ei{2}}(w_1,w_2;\sigma) & = \frac{\varphi(w_2;\sigma)^{-1}}{u_2-u_1} \Big|_{\sigma_i=\sigma} C_{\ei{1}\ei{2}} \ , \qquad \qquad \varphi(w;\sigma) = \frac{(1+b\,\s) (1-u^2)}{u^2} \ . \la{car}
\end{split}\end{equation}
Let us see what happens if we naively use the charges defined in eq.~\eqref{charges}. For $n=2$ we get
\begin{equation}
Q^{(2)}_{\pm1} = \operatorname{res}_{u = \pm 1} \int d\sigma \, \Tr\big[\varphi(w;\sigma)\widehat{L}(w;\sigma)^2\big]
= \mp \tfrac{1}{2}\int d\sigma \, \Tr\big[(1+b\,\s) j \pm 2 j X + \frac{1}{1+b\,\s} X^2\big] \ .
\end{equation}
We immediately see that, just as for the standard PCM ($b=0$), the Hamiltonian and the spatial momentum are given by
\begin{equation}\begin{split}
\widehat{\mathcal{H}} &=\ha ( Q^{(2)}_{+ 1} - Q^{(2)}_{-1})
= - \tfrac{1}{2} \int d\sigma \, \Tr\big[ \FF \, j^2 + \FF^{-1} X^2 \big] \ , \qquad \qquad \FF\equiv 1+b\,\s \ , \\
\widehat{\mathcal{P}} &=\ha ( Q^{(2)}_{+ 1} + Q^{(2)}_{-1})
= - \int d\sigma \, \Tr\big[jX\big] \ .\la{hpp}
\end{split}\end{equation}
The Hamiltonian, which governs the time evolution of the system, is conserved since it does not have any explicit $\tau$ dependence.
On the other hand, the spatial momentum is not conserved for $b\neq 0$
\begin{align} \frac{d}{d\tau} \widehat{\mathcal{P}}&=
\partial_\tau \widehat{\mathcal{P}} - \{\widehat{\mathcal{H}} , \widehat{\mathcal{P}}\} \no\\
&= -
\tfrac12\int d\sigma \, \partial_\sigma\Tr\big[(1+b\,\s)j^2 + \frac{1}{1+b\,\s} X^2\big] - \tfrac{b}{2} \int d\sigma \, \Tr\big[j^2 - \frac{X^2}{(1+b\,\s)^2} \big] \ ,
\end{align}
reflecting the explicit dependence of the Lagrangian \eqref{eq:lagsig} on $\sigma$.
Turning now to the cubic and quartic charges, one can check that there is no linear combination of these that is conserved for $b\neq 0$, and we expect this to be the case for all $n > 2$.
Therefore, this naive use of eq.~\eqref{charges} does not allow us to construct local conserved charges when the coupling is space dependent.

It turns out we can do slightly better, and to discuss this we turn to the case of the model \rf{lac} with the general local coupling
admitting a Lax connection
(setting $f^\pm(\xi^\pm) = \tfrac12(1+b \hat{f}^\pm(\xi^\pm))$, \ $\xi^\pm = \ha( \t\pm \s)$ and $c=1$),
\begin{equation}\la{hah}
\widehat{\mathcal{L}} = -\tfrac{1}{2}\, h(\xi^+,\xi^-)\, \Tr[J_+J_-] \ ,\qquad \qquad h \equiv 1+\tfrac12 b \big[\hat{f}^+(\xi^+) + \hat{f}^- (\xi^-)\big] \ .
\end{equation}
Defining momenta as in \eqref{eq:pi}, we now have from \rf{jX} that $X =
h\, g^{-1} \partial_\tau g$ and the Lax matrix is
(cf. \rf{lac},\rf{laj})
\begin{equation}
\widehat{L}(w;\sigma) = \frac{ z + z^{-1}+2}{4}\, j + \frac{z - z^{-1}}{4 h }
\, X \ ,\qquad \qquad z(w,\sigma) = \sqrt{\frac{w -1 + b \hat{f}^-}{w+1+b\hat{f}^+}} \ .
\end{equation}
Once again the Lax matrix algebra can be written in the form \eqref{laxalg} with the dynamical $R$-matrix given by (cf. \rf{rmat})
\begin{equation}
R_{\ei{1}\ei{2}}(w_1,w_2;\sigma) = \frac{(1+z_1)(1+z_2)(1-z_2-z_2^{-1})}{8(z_1-z_2) \, h
}\Big|_{\sigma_i=\sigma}
C_{\ei{1}\ei{2}} \ ,
\end{equation}
which again solves the cYBE \eqref{eq:cybe}.
As before, introducing
$u(w;\sigma) = \frac{\sqrt{w-1-b\hat{f}^-}-\sqrt{w+1+b\hat{f}^+}}{\sqrt{w-1-b\hat{f}^-}+\sqrt{w+1+b\hat{f}^+}}$ and $u_i = u(w_i;\sigma_i)$,
the Lax matrix and $R$-matrix may be written in the form
\begin{equation}\begin{split}
\widehat{L}(w;\sigma) & = \frac{1}{1-u^2}j+
h^{-1} \frac{u}{1-u^2} X \ ,
\\
\mathcal{R}_{\ei{1}\ei{2}}(w_1,w_2;\sigma) & = \frac{\varphi(w_2;\sigma)^{-1}}{u_2-u_1} \Big|_{\sigma_i=\sigma} C_{\ei{1}\ei{2}} \ , \qquad \varphi(w;\sigma) = \frac{h (1-u^2)}{u^2} \ .
\end{split}\end{equation}
Let us
again naively consider the charges defined in eq.~\eqref{charges}.
We find that the corresponding
Hamiltonian $\widehat{\mathcal{H}}$ and spatial momentum $\widehat{\mathcal{P}}$ are
given by the same expressions as in \rf{hpp}, now with $h$ given in \rf{hah}.
The time evolution of $\widehat{\mathcal{H}}$ and $\widehat{\mathcal{P}}$ is given by
\begin{equation}\begin{split}\label{eqeq}
\partial_\tau \widehat{\mathcal{H}} - \{\widehat{\mathcal{H}},\widehat{\mathcal{H}}\} &
= - \tfrac{b}{8} \int d\sigma \, (\hat{f}^+{}' + \hat{f}^-{}')\Tr\big[j^2 - \FF^{-2} X^2\big] \ ,
\\
\partial_\tau \widehat{\mathcal{P}} - \{\widehat{\mathcal{H}},\widehat{\mathcal{P}}\} & =
- \tfrac{1}{2} \int d\sigma \, \partial_\sigma \Tr\big[\FF j^2 + \FF^{-1}X^2\big]
- \tfrac{b}{8} \int d\sigma \, (\hat{f}^+{}' - \hat{f}^-{}')\Tr\big[j^2 - \FF^{-2} X^2\big] \ .
\end{split}\end{equation}
For general functions $\hat{f}^\pm$,
we see that
$\widehat{\mathcal{H}}$ and $\widehat{\mathcal{P}}$ are not conserved.
However, if there exist constants $c_{ \mathcal{H} }$ and $c_{ \mathcal{P} }$ such that
$c_{ \mathcal{H} }(\hat{f}^+{}' + \hat{f}^-{}') + c_{ \mathcal{P} } (\hat{f}^+{}' - \hat{f}^-{}') = 0$, then the linear combination $c_{ \mathcal{H} } \widehat{\mathcal{H}} + c_{ \mathcal{P} } \widehat{\mathcal{P}}$ of quadratic charges
is conserved.
Recall that we consider the theory on an infinite spatial interval and assume decaying boundary conditions such that the boundary term in \eqref{eqeq} vanishes.
There are two classes of solutions to this condition: (i) $\hat{f}^\pm(\xi^\pm) = \frac{\pm\xi^\pm}{c_{ \mathcal{H} } \pm c_{ \mathcal{P} }}$, and (ii) $c_{ \mathcal{H} } = c_{ \mathcal{P} }$, $\hat{f}^+(\xi^+) = \text{const}$
or $c_{ \mathcal{H} } = - c_{ \mathcal{P} }$, $\hat{f}^-(\xi^-) = \text{const}$.
For the first class the coupling $\FF$ in \rf{hah}
is linear in $\tau$ or $\sigma$ and the conserved charge comes from the unbroken translation symmetry.
The second class gives the `chiral' theories discussed in Appendix \ref{chiral}.

The cubic charges that follow from eq.~\eqref{charges} are
\begin{equation}\begin{split}
\widehat{\mathcal{H}}^{(3)} & = \ha ( Q_{+1}^{(3)} - Q_{-1}^{(3)}) =
-\tfrac{1}{2}\int d\sigma \, \Tr\big[\FF^k j^3+ 3 \FF^{k-2} j X^2 \big] \ ,
\\
\widehat{\mathcal{P}}^{(3)} & =\ha ( Q_{+1}^{(3)} + Q_{-1}^{(3)}) =
-\tfrac{1}{2}\int d\sigma \, \Tr\big[3 \FF^{m+2} j^2 X + \FF^{m} X^3 \big] \ ,
\end{split}\end{equation}
with $k=2$ and $m=-1$.
Leaving $k$ and $m$ arbitrary we find that their time evolution is given by
\begin{equation}\begin{split}
\partial_\tau \widehat{\mathcal{H}}^{(3)} - \{\widehat{\mathcal{H}},\widehat{\mathcal{H}}^{(3)}\} &
= - \tfrac{1}{2} \int d\sigma \, \partial_\sigma \Tr\big[3\FF^{k-1}j^2 X + \FF^{k-3} X^3\big]
\\ & \quad- \tfrac{b}{8} \int d\sigma \, \Big((\hat{f}^+{}'+\hat{f}^-{}')\Tr\big[k \FF^{k-1} j^3 + 3(k-2)\FF^{k-3} jX^2\big]
\\ & \hspace{60pt} -(\hat{f}^+{}'-\hat{f}^-{}')\Tr\big[3(k-2)\FF^{k-2}j^2 X + k \FF^{k-4} X^3\big] \Big)\ ,
\\
\partial_\tau \widehat{\mathcal{P}}^{(3)} - \{\widehat{\mathcal{H}},\widehat{\mathcal{P}}^{(3)}\} &
= - \tfrac{1}{2} \int d\sigma \, \partial_\sigma \Tr\big[3\FF^{m+1}X^2j + \FF^{m+3}j^3\big]
\\ & \quad- \tfrac{b}{8} \int d\sigma \, \Big((\hat{f}^+{}'+\hat{f}^-{}')\Tr\big[3(m+2)\FF^{m+1} j^2 X + m \FF^{m-1} X^3\big]
\\ & \hspace{60pt} -(\hat{f}^+{}'-\hat{f}^-{}')\Tr\big[m \FF^{m+2} j^3 + 3(m+2)\FF^m j X^2 \big] \Big)\ .
\end{split}\end{equation}
From these expressions we find that there is no linear combination of these charges that is conserved for $k=2$ and $m=-1$, i.e.\ the values of $k$ and $m$ that follow from eq.~\eqref{charges}.
Nevertheless, we may ask if
such a charge can be constructed
by modifying the values of $k$ and $m$.
One can show that this is only possible for case (ii) above, i.e.\ for the `chiral' theories, and requires us to take $k=-m = \frac{3}{2}$.
For these values $\widehat{\mathcal{H}} \pm \widehat{\mathcal{P}}$ is conserved for $\hat{f}^\pm(\xi^\pm) = \text{const}$.
It is natural to expect that for these `chiral' theories a similar construction should hold for all $n$, with the conserved charges corresponding to the holomorphic conserved currents derived in Appendix \ref{chiral}.
\unskip\footnote{We can also consider more general ansatze for the quadratic and cubic charges
\begin{equation*}
\widetilde{\mathcal{Q}}^{(n)}
= - \tfrac{1}{2} \int d\sigma \, \Tr\big[ \sum_{i=0}^n \mu_{i,n} \FF^{\frac{n}{2}-i} j^{n-i}X^i \big] \ , \qquad \FF = f^+(\xi^+) + f^-(\xi^-) \ ,
\end{equation*}
where $\mu_{i,n}$ are arbitrary functions of $(\tau,\sigma)$.
Analysing the time evolution of $\widetilde{\mathcal{Q}}^{(2)}$ and demanding it is conserved, we find that $\mu_{0,2} = \mu_{2,2} = \nu_2^+ + \nu_2^-$ and $\tfrac12\mu_{1,2} = \nu_2^+ - \nu_2^-$, with the functions $\nu_2^\pm = \nu_2^\pm(\xi^\pm)$ subject to $\nu_2^+f^+{}' = \nu_2^-f^-{}'$.
For general $f^\pm{}' \neq 0$ we can solve this equation to construct one conserved charge.
For $n=3$ we find that $\mu_{0,3} = \tfrac13\mu_{2,3} = \nu_3^+ + \nu_3^-$ and $\mu_{3,3} = \tfrac13\mu_{1,3} = \nu_3^+ - \nu_3^-$ with $\nu_3^\pm = \nu_3^\pm(\xi^\pm)$ subject to $\nu_3^+f^+{}' = \nu_3^-f^-{}' = 0$.
For $f^\pm{}' \neq 0$ it follows that $\nu_3^\pm = 0$ and we still do not find a cubic conserved charge.

For the `chiral' theories without loss of generality let us take $f^+{}' = 0$.
It follows that $\nu_{2,3}^- = 0$ while $\nu_{2,3}^+$ is a free function and we have both quadratic and cubic charges as expected.
For the case of constant coupling we have that both $\nu_{2,3}^+$ and $\nu_{2,3}^-$ are free functions.}

We have seen that, even in the case of the `chiral' theories, the naive application of
the standard expression \eqref{charges} for the local charges
above needs to be modified.
This suggests
the need for
a more systematic attempt to construct an infinite tower of local conserved charges
-- and
understand the underlying algebraic structures --
to determine the status of integrability
in these models with local couplings.

\section{Discussion}\la{s6}

In this paper we
observed
a surprising new connection between
classical integrability and the RG flow in 2d theories:
starting with an integrable theory and promoting its couplings to time-dependent functions, its
Lax pair generalizes naturally to the resulting
time-dependent theory only if the coupling functions
solve the 1-loop RG equations
of the original theory.
We demonstrated this on six classes of integrable $\s$-models.

One interesting implication is
that the 1-loop $\beta$-functions, which are normally associated with 1-loop divergences in quantum theory,
can thus be obtained in these models through the classical procedure of requiring the existence of a Lax pair in
the corresponding time-dependent theory.

As we discussed in the Introduction,
such time-dependent models can be naturally embedded into string theory by
starting with a Weyl-invariant $\s$-model \rf{csm} with two extra `null' directions $(u,v)$
and fixing a l.c.\ gauge $u=\t$.
If the l.c.\ theory \rf{lct} is integrable, it is natural to expect that the corresponding
string theory should be solvable.\foot{The classical solvability of the l.c.\ theory \rf{lct} should be equivalent to solvability of the full theory \rf{csm} since one may move
from $u=\t$ to
any other
solution of the $u$-equation $\del_+\del_- u = 0 $
by a conformal transformation. The $v$-equations
$ \del_\pm v =\ha G_{ij}(\t,x) \, \del_\pm x^i \del_\pm x^j$
(following from the conformal constraints)
are
linear and so are also readily solvable. It is not immediately clear, however,
if integrability in the sense of admitting a Lax representation should similarly lift up to the $(D+2)$-dimensional string $\s$-model.}

The Lax connections for the time-dependent theories have an
unusual form (depending explicitly on $\t$ and $\s$ as in the special
models discussed in \cite{Belinsky:1971nt,Breitenlohner:1986um})
but, given certain boundary conditions, the entries or
the eigenvalues
of the monodromy matrix on an infinite spatial line are conserved.
Due to the explicit $\s$ dependence in the Lax connection, it is hard to
evaluate the monodromy and find the conserved charges explicitly.
We considered some consistent reductions of the 2d equations of motion to time-dependent 1d systems.
The `trivial' ($\s$-independent)
reduction of the PCM is clearly integrable due to the remaining global symmetry.
For the `non-trivial' reduction in section \ref{s43} we evaluated the monodromy, and hence constructed a conserved charge \rf{cq}, in perturbation theory in the small field expansion. The perturbative existence of this conserved charge (following from the existence of a Lax connection)
is a non-trivial property, suggesting
that the 1d theory \rf{s1} may be interpreted as integrable.\foot{The time-dependent 1d theory \rf{s1} is certainly not integrable in the Liouville sense as it would fail the Kovacic algorithm test used in \cite{Basu:2011fw,Stepanchuk:2012xi}.
Here we are suggesting a broader notion of integrability relating it to the explicit solvability of the equations of motion.
Note that, adding a generic (time-dependent) potential to the linearized theory \rf{lt}, we would not expect the linearized conserved charge \rf{ql} to admit an extension to a conserved charge of a non-linear theory.}

One way to fully establish the classical integrability of the $\s$-models with local couplings
would be to construct an infinite tower of local conserved charges in involution, in the spirit of \cite{Goldschmidt:1980wq,Evans:1999mj,Evans:2000qx,Lacroix:2017isl}.
In section \ref{s5} we attempted this
for the PCM
in the Hamiltonian formulation.
A naive application of the method used for constant couplings does not yield such charges, except in the case of the `chiral' theories
where it works with a slight modification.
Nevertheless, the form of the Lax matrix and $R$-matrix is suggestive of an underlying algebraic structure.
Understanding it may provide additional insights into the question of integrability.

While we considered time-dependent models at the classical level only, one may wonder if they themselves
are renormalizable, i.e.\ stable under RG flow. Renormalization of generalized
$\s$-models with target-space metric depending on 2d coordinates was considered in \cite{Osborn:1987au}. In the case
we discussed (cf.\ \rf{lct}) when $G_{ij}$ depends only on $\t$, i.e. $\L=G_{ij}(\t, x) \del_+ x^i \del_- x^j$, the 1-loop logarithmic counterterm
is proportional to $ K_1= G_{ml} G^{jk} \del_\t x^m\, \del_\t \Gamma^l_{jk} + \frac{1}{4} \del_\t G^{ij} \del_\t G_{ij} $, where
$\Gamma^l_{jk}$ is the Christoffel connection of $G_{ij}$. Thus in general one needs to add also counterterms with one and no derivatives, i.e. $V_m(x) \del_\t x^m + T(x)$. However, in the `factorized' case when $G_{ij} (\t, x) = f(\t) G_{ij}(x),$
like in the time-dependent PCM \rf{lcPCM}, the above counterterm becomes an $x$-independent function,
$K_1\sim ( \del_\t f)^2 $. Furthermore, if $f$ is linear in $\tau$ as in \rf{lcPCM} then this $K_1$ is just a constant
and thus such a model is renormalizable.
This suggests that at least some time-dependent integrable models
discussed in this paper may have
well-defined quantum generalizations.

\medskip

Among other possible directions, it would be important to generalize the present investigation to
other integrable $\sigma$-models, e.g.\ the class of models based on complex homogeneous spaces
whose  Ricci flow was studied in \ci{byk} and  the $\mathbb{Z}_m$ coset models of \cite{yo}.
Furthermore, while in this paper we have only studied 2-derivative $\s$-models,
we expect the connection between
the requirement of integrability of the time-dependent theory and RG flow to be more general and to apply also to
models with potentials.
Indeed,
one also finds this remarkable connection in the case of the sine-Gordon model,
\be
S= \tfrac{1}{4\pi} \int d\t d\s \, \tfrac{1}{g^2}\big[ \ha \del_+ x \del_- x + m^2 \cos{x}\big]\ . \la{sgm}
\ee
Replacing the couplings $(g,m)$ by functions\foot{For a discussion of the sine-Gordon model with a particular `non-integrable'
local coupling $m=m(\s)$ see \cite{Levkov:2020lfa} and references there.}
of $\t$
one can show
that, for the resulting theory to be integrable
(assuming a natural ansatz for a generalization of the Lax connection to the time-dependent theory
--
see Appendix \ref{sing}),
the functions $(g(\t),m(\t))$ should be solutions of the 1-loop RG flow equation for \rf{sgm}, i.e.
should be given by
\be m^2(\t) = e^{ \b(g)\, \t} \, m_0^2 \ , \ \ \ \qquad g(\t) = g \ , \ \ \ \ \qquad \b(g) = - 2 + g^2 \ . \la{52}\ee
The time-dependent theory,
\be
\widehat{S}= \tfrac{1}{4\pi} \int d\t d\s \, \tfrac{1}{g^2} \big[ \ha \del_+ x \del_- x + e^{\b(g)\, \t}\, m^2_0\, \cos{x}\big]\ ,\la{53}
\ee
is indeed classically integrable since the explicit $\t$-dependence in \rf{53} can
be removed by a 2d conformal transformation getting back to \rf{sgm}:
under $\xi^\pm \to f^\pm (\xi^\pm)$ the action
\rf{sgm} retains its form with
$m^2 \to {f^+}'(\xi^+)\, {f^-}'(\xi^-)\, m^2 $.
Note that
the non-trivial 1d reduction of the $SU(2)$ PCM in eq.\ \rf{s1} is also obtained as the `trivial' reduction $x=x(\t)$
of \rf{53}
after a redefinition $\t \to \log \t $, supporting the expectation that \rf{s1} is an integrable 1d theory.
It would be interesting to explore
whether
this time-dependent integrability--RG flow connection applies also to
other examples of integrable massive theories, such as complex sine-Gordon and Toda models.

\medskip

One of the features of our construction of the Lax connection for time-dependent theories is that
the constant spectral parameter of the original theory
is replaced by
a function
$z\to z(w;\t,\s)$ of a new spectral parameter $w$ and the 2d coordinates (cf.
\rf{pla},\rf{lpl}).
This puts the spectral parameter and 2d space-time coordinates on a more equal footing,
suggesting a possible interpretation from the point of view of the
construction \cite{Costello:2019tri}
of many
integrable 2d theories from a 4d Chern-Simons theory,
with the two extra directions related to the complex spectral parameter (see also \cite{Delduc:2019whp} and refs.\ there).
In that context the redefinition of $z$ is like changing
the differential structure of the 4d space, i.e.\ replacing $\del_{\t,\s} z = 0$
with $\del_{\t,\s} z = V_{\t,\s}(z;\t,\s)$ (cf.\ \rf{pd1}).
It would be interesting to see if the time-dependent models considered above
can indeed be reproduced from the 4d Chern-Simons construction.

\medskip

\section*{Acknowledgments}

We would like to thank G. Arutyunov, S. Lacroix, A. Polyanin and B. Vicedo for useful discussions
and comments on the draft.
We are also grateful to T. McLoughlin for an important comment on the relation to earlier work.
BH was supported by the Swiss National Science Foundation through the NCCR SwissMAP.
NL was supported by the EPSRC grant EP/N509486/1.
AAT was supported by the STFC grants ST/P000762/1 and ST/T000791/1.

\bigskip

\appendix

\section{Details of derivation of RG flow from existence of Lax connection}\label{A}
\def\theequation{A.\arabic{equation}}
\setcounter{equation}{0}

Here we shall provide some further details of the derivation of the RG flow
from the consistency of the Lax representation for the $(\t,\s)$-dependent model in Section 3.

\addtocontents{toc}{\protect\setcounter{tocdepth}{1}}
\subsection{Derivation of the equations \rf{alg},\rf{als}}\la{det}

To derive \rf{alg},\rf{als}, we shall ignore terms proportional to the derivatives of the couplings $\del_\pm h_\a$ and match the other terms between the flatness condition of the Lax connection and the equation of motion. The algebraic equations \rf{alg},\rf{als} follow essentially because these `non-derivative'
terms do not change upon the introduction of space-time dependence of the couplings.

Let us consider
the cases of models associated to a group and to a symmetric space
separately.
For the group space case, the original equations of motion following from the flatness of Lax in \rf{gl} are the flatness
and the conservation of the current $\mc{A}_\pm$.
In the $(\t,\s)$-dependent model \rf{sth}, these equations may only be modified by $\O(\del h)$ terms
\be
F_{+-}(\mc{A})= \O(\del h) \ , \qquad\qquad \del_+ \mc{A}_- + \del_- \mc{A}_+ = \O(\del h) \ . \la{gc1}
\ee
They should follow from the flatness of the Lax connection ansatz \rf{glt},
\be
F_{+-} (\widehat{L}) = p_+ p_- F_{+-} (\mc{A}) + \big[p_-(1 - p_+) \del_+ \mc{A}_- - p_+(1-p_-) \del_- \mc{A}_+\big]+ \O(\del p) \ . \la{gc2}
\ee
Comparing first
the terms with $\del_\pm \mc{A}$ and $\mc{A}^2$, i.e.\ neglecting the $\O(\del h)$ and $\O(\del p)$ terms in \rf{gc1},\rf{gc2},
the matching of \rf{gc1} and \rf{gc2} then implies
\be
p_-(1 - p_+) = - p_+(1-p_-) \ .
\ee
This leads to the algebraic equation \rf{alg}.

For the symmetric space case, the original equations of motion following from the vanishing of the curvature of \rf{sl} are the flatness condition for $(\mc{B}+\mc{P})_\pm$ and the equations $D^{\mc{B}}_\pm \mc{P}_\mp = 0$, where $D^{\mc{B}}_a$ is the covariant derivative with respect to the connection $\mc{B}_a$ in the subalgebra $\Lie (H)$.
In the $(\t,\s)$-dependent model \rf{sth} these equations are modified, as in \rf{gc1},
\be
F_{+-}(\mc{B+P})= \O(\del h) \ , \qquad D^\mc{B}_\pm \mc{P}_\mp = \O(\del h) \ . \la{sc1}
\ee
These should follow from the flatness of the ansatz for the Lax pair in \rf{slt}
\begin{align}
F_{+-} (\widehat{L}) ={} &q_+ q_- F_{+-}(\mc{B+P}) + (r_- - q_+ q_-) D^\mc{B}_+ \mc{P}_- - (r_+ - q_+ q_-) D_-^\mc{B} \mc{P}_+
\no \\
& \ + q_-(1-q_+)\del_+ \mc{B}_- - q_+(1-q_-)\del_- \mc{B}_+ + (r_+ r_- - q_+ q_-)[\mc{P}_+, \mc{P}_-] \no \\
& \ + r_- (q_+ -1)[ \mc{B}_+, \mc{P}_-] - r_+ (q_- -1) [ \mc{B}_-, \mc{P}_+]
+ \O(\del q,\del r) \ . \la{sc2}
\end{align}
Comparing the terms in \rf{sc1} and \rf{sc2} that contain
$\del \mc{B}$, $\del \mc{P}$, $\mc{B}^2$, $\mc{P}^2$ and $\mc{BP}$, i.e.\ neglecting the $\O(\del h)$ terms in \rf{sc1} and $\O(\del q, \del r)$ terms in \rf{sc2},
we conclude in particular that the coefficients of the extra $\del\mc{B}$, $[\mc{B,P}]$ and $[\mc{P,P}]$ terms in $\rf{sc2}$ must vanish. Assuming $q_\pm,r_\pm$ are not all zero (so that the Lax connection \rf{slt} is not identically zero) the only solution is $q_+ = q_- = 1$ and $r_+ r_- = 1$, i.e.\ the conditions in \rf{als}.

\subsection{RG flow in PCM case}\la{fdr}

The matching of the $\O(\del h)$ terms then forces the coupling functions to solve the RG flow equations. The general structure of this argument is explained in eqs. \rf{pd1}-\rf{RGs2}, but here we shall run through it explicitly for the simplest PCM example.

Together, \rf{glt} and \rf{alg} lead to the following ansatz for the Lax connection (cf.\ \rf{lpl})
\be
\widehat{L}_\pm = \ha \Big(1+[z(\t,\s)]^{\pm 1}\Big) J_\pm \ , \la{laxh}
\ee
whose curvature is ($F_{+-}(J)=0$ since $J=g^{-1} dg$)
\begin{align}
F_{+-}(\widehat{L}) &= \tfrac{1-{z}^2}{	4{z}} (\del_+ J_- + \del_- J_+) - \tfrac{\del_+ z}{2z^2} J_- - \tfrac{\del_- z}{2} J_+
\ . \la{lah}
\end{align}
The equation of motion for the PCM with coupling $h(\t,\s)$ is (generalizing \rf{pe1})
\be
h (\del_+ J_- + \del_- J_+) + \del_+ h J_- + \del_- h J_+ = 0 \ . \la{eoh}
\ee
For this to be equivalent to the vanishing of the curvature \rf{lah}
the ratios
of the different
coefficients should match.
This leads to the equations of the form \rf{pd1}, i.e.
\be\la{a9}
\del_+ z = - \tfrac{z}{2} (1-z)^2 \,\tfrac{\del_+ h}{h} \ , \qquad \del_- z = - \tfrac{1}{2z} (1-z)^2 \,\tfrac{\del_- h}{h}\ .
\ee
The consistency condition $\del_+(\del_- z) - \del_- (\del_+ z) = 0$ gives
\be
\tfrac{(1-z^2)^2}{2h z} \ \del_+ \del_- h = 0 \ .
\ee
It is remarkable that the $z$ dependence has totally factored out
(a term proportional to $\del_+ h \del_- h$ is absent due to a special cancellation).
Excluding the trivial cases $z(\t,\s) =\pm 1 $, which would not encode the correct equation of motion, we find that the Lax connection \rf{laxh} only matches the correct equation of motion if $\del_+ \del_- h =0$, i.e.\ if
$h = f^+ (\xi^+) + f^-(\xi^-)$.
Any such solution is related to the 1-loop RG flow $h=c\, \t$ by a 2d conformal transformation.

\subsection{RG flow for theories with multiple couplings}\la{mr}

In section \ref{s3} (see eqs. \rf{pd1}-\rf{RGs2}) we explained the derivation of the RG flow
focussing on the case with only one coupling.
The same conclusion also holds for the group space $\eta$-model with two couplings $h$ and $\eta$ (see Table \ref{tab1}), and more generally is expected to be true for multi-coupling theories.

There are multiple independent structures in the equation of motion (for the $\eta$-model these involve different
powers of the $R$-matrix) and correspondingly in the flatness of the Lax connection. Matching
the coefficients of these structures in the $\eta$-model case yields \textit{two} pairs of equations (cf.\ \rf{pd1}),
\begin{align}
&\del_\t z = U_\t(z;h,\eta) \ , \qquad \del_\s z = U_\s(z;h,\eta) \ , \la{sy1}\\
&\del_\t z = V_\t(z;h,\eta) \ , \qquad \del_\s z = V_\s(z;h,\eta) \ . \la{sy2}
\end{align}
In general, for an $N$-coupling theory we would expect to find $N$ pairs of equations.

As a system of equations for $z$, \rf{sy1},\rf{sy2} is clearly overdetermined. In two combinations of these equations the $z$ dependence cancels to give relations between $h(\t,\s)$ and $\eta(\t,\s)$
\begin{align} \la{a13}
\del_\t( \eta h^{-1} ) = \del_\s ( \eta h^{-1} ) = 0 \ .
\end{align}
Eq.\ \rf{a13}
implies that $\nu \equiv \eta h^{-1}$ is a constant; this coincides precisely with the 1-loop RG invariant
of the $\eta$-model (see Table \ref{tab1}).
For an $N$-coupling theory, we may expect to obtain $(N-1)$ RG invariants $\nu_r$ in this way.

Then
the system \rf{sy1},\rf{sy2} reduces to just two equations -- effectively returning to the single-coupling case of equations \rf{pd1}. Again, the consistency condition for the two remaining equations takes the remarkable form \rf{2w}, where the beta function $\b(h) \equiv \beta^h (h, \nu )$ is understood as a function of the coupling $h$ and the RG invariant. As in eqs.\ \rf{cfs},\rf{RGs2} it then follows (modulo a conformal transformation) that $h(\t,\s)$ depends only on $\t$ and follows the 1-loop RG flow.
The same should generalize to the $N$-coupling case with
$(N-1)$ independent RG invariants $\nu_r$ that can be chosen as constants.

\section{On non-local charges in time-dependent symmetric space \texorpdfstring{$\l$}{λ}-model}\label{om}
\def\theequation{B.\arabic{equation}}
\setcounter{equation}{0}

In Section 4 we discussed the construction of non-local charges in the time-dependent PCM.
Here we shall comment on the other models in Table \ref{tab1},
and, in particular, on the symmetric space $\l$-model.

The construction of the conserved monodromy matrix \rf{p1},\rf{px} works similarly, although for the models built on symmetric spaces (symmetric space $\s$-model and symmetric space $\l$-model), it is only the eigenvalues of the monodromy matrix that are conserved. A sufficient boundary condition in all cases is that $g(\t,\s) \to g_0$ at spatial infinity and that $(g-g_0)$ decays sufficiently fast so
that the monodromy converges at spatial infinity. As in the PCM example it is hard to evaluate the conserved charges explicitly and thus to verify that they are infinite in number (i.e.\ depend non-trivially on the spectral parameter $w$).

Except for the symmetric space $\l$-model, all the other models have global symmetries.
As for the PCM in \rf{noet}, the associated
charges can be obtained by expanding the monodromy around $w=\infty$.\foot{This applies also for the symmetric space $\s$-model after applying the gauge transformation $L_\pm \to g L_\pm g^{-1} - \del_\pm g g^{-1}$ to the Lax connection $L = J^H_\pm + z^{\pm 1} J_\pm^{G/H}$, obtaining the alternate Lax connection $L_\pm = \ha (1-z^{\pm 1}) (-2 g J_\pm^{G/H} g^{-1} )$ of the `group space' form \rf{gl}, instead of \rf{sl} (modulo sign reversal of $z$).}
In the remainder of this appendix, we shall consider
the trivial reduction of the time-dependent symmetric space $\l$-model, which has no manifest global symmetries.
In general, the eigenvalues of the monodromy matrix would be conserved on an infinite spatial line;
however, in the trivial reduction the monodromy matrix does not converge at spatial infinity since $L_\s$ does not vanish there.
Below we will try
to shed some light on this issue.

Let us first recall what happens for geodesics in the usual time-independent case, where the Lax connection is $L_\pm = A_\pm^H + z^{\pm 1} \tfrac{1}{\sqrt{\l}} A_\pm^{G/H}$. In the trivial reduction we have $A_\pm = A_\pm (g(\t))$ so
the periodicity condition \rf{pe} is satisfied and the eigenvalues of the monodromy conserved even on a finite interval of length $a $. The path-ordered exponential trivializes to give $\M = \exp{ (a \, L_\s )}$ and hence, equivalently,
the eigenvalues of $L_\s = \ha (A_+^H - A_-^H + \tfrac{z}{\sqrt{\l}} A_+^{G/H} - \tfrac{z^{-1}}{\sqrt{\l}} A_-^{G/H})$ are conserved. For example,
let us consider the simplest $SO(3)/SO(2)$ $\l$-model (with the subgroup $SO(2)$ generated by $\s_1$), parametrized after gauge fixing as
\begin{align}
&g = e^{i \a \s_3} e^{i \b \s_1} \ , \qquad\qquad \cos{\a} = \sqrt{p^2+q^2} \ , \qquad \ \cos{\a}\cos{\b} = p \ , \\
&\L = \frac{k}{1-\l^2} \frac{(1+\l)^2 (\del p)^2 + (1-\l)^2 (\del q)^2}{1-p^2-q^2} \ . \la{2dl}
\end{align}
There is only one independent eigenvalue $l$ of $L_\s$ since it is a traceless $2\times 2$ matrix. Its expansion around, e.g.,
$z=1$ gives (at least) two independent conserved charges for the geodesic motion
\begin{align}
&\dot{Q}_1 = \dot{Q}_2 = 0 \ , \qquad \qquad l = \tfrac{1}{2(\l^2-1)}\sqrt{Q_1} + (z-1)^2 \tfrac{\l}{\l^2-1} \frac{Q_2}{\sqrt{Q_1}} + \ldots \ ,\\
&\quad Q_1 = \frac{(1+\l)^2[4\l(p^2-1)-(\l-1)^2q^2]\dot{p}^2 + 2(\l^2-1)^2 p q \dot{p}\dot{q} - (\l-1)^4p^2\dot{q}^2}{(1-p^2-q^2)^2} \ , \la{q1} \\
&\quad Q_2 = \frac{(1+\l)^2\dot{p}^2 + (1-\l)^2 \dot{q}^2}{1-p^2-q^2} \ , \la{q2}
\end{align}
with $Q_2$ proportional to the Hamiltonian.

Returning to the time-dependent case, the monodromy is not defined on an infinite line, and on any finite interval the periodicity condition is not satisfied due to the explicit $\s$ dependence of the Lax connection for generic values of $w$. However, at certain special values of $w$ satisfying $\exp{(\tfrac{c}{k}w)}=\pm \infty$, the $\s$ dependence disappears to give flat connections
\begin{align}
&[\del_+ + A_+^H + \tfrac{1}{\l(\t)} A_+^{G/H} , \ \del_- + A_-^H + A_-^{G/H} ]=0 \ , \la{f1} \\
&[\del_+ + A_+^H + A_+^{G/H} , \ \del_- + A_-^H + \tfrac{1}{\l(\t)} A_-^{G/H} ] = 0 \ , \qquad \qquad \l(\t) = \exp{(\tfrac{c}{k} \t)} \ , \la{f2}
\end{align}
generalizing the same expressions from the time-independent case ($\l(\t) \to \l$). At these values the periodicity condition \textit{is} satisfied on any finite interval. The monodromy trivializes as in the time-independent case to give $\M = \exp{(a\, L_\s)}$, so the eigenvalues of $L_\s$ are again conserved.

In fact, the two flat connections \rf{f1},\rf{f2} are related by a gauge transformation, so their conserved charges are the same. Hence, the maximum number of independent charges obtained from the flat connections \rf{f1},\rf{f2} is $r = {\rm rank}(G)$
for a symmetric space $G/H$. This number is generally less than the number of fields, $\dim{G/H} =\dim{G}-\dim{H}$ (e.g.\ for $SO(n+1)/SO(n)$ we get $r= n-1 < \dim{SO(n+1)/SO(n)} = n$), so this is not sufficient for integrability.

For example, in the $SO(3)/SO(2)$ case \rf{2dl}, where there are 2 fields, we only obtain $r=1$ conserved charge
\be
Q = \frac{[\l(\t)+1]^4 (p^2-1) \dot{p}^2 + 2[\l(\t)^2-1]^2 p q \dot{p}\dot{q}+ [\l(\t)-1]^4(q^2-1)\dot{q}^2}{[\l(\t)^2-1]^2(1-p^2-q^2)^2} \ .
\ee
In the time-independent limit $\l(\t) \to \l$ (obtained, e.g.,
by shifting $\t \to \tfrac{k}{c}\log{\l} + b \t$ and taking $b\to 0$), this charge becomes a particular combination of the charges \rf{q1},\rf{q2} in the time-independent theory,
\be
Q \to Q_1 - (\l-1)^2 Q_2 \ .
\ee
Having restricted consideration to special values of $w$, we do not find
enough charges for the integrability of the geodesics.
Since the construction of charges is subtle, depending on boundary conditions and the choice of the spatial domain
(the periodicity condition \rf{pe} must be satisfied and, if the interval is infinite, the monodromy must converge at infinity),
it is possible that the monodromy constructed on some special domain would yield further conserved charges,
but this remains to be clarified.

\section{Time-dependent 1d harmonic oscillator and conserved charge}\label{C}
\def\theequation{C.\arabic{equation}}
\setcounter{equation}{0}

In section \ref{s4} we came across a particular time-dependent linear model \rf{lt}.
Starting with a general time-dependent linear 1d action\foot{We do not include the term $f(\t) \theta \dot{\theta}$
as it can be put in the form $k(\t) \theta^2$ by adding a total derivative.}
\be
S = \int d\t \, \big[ h(\t)\, \dot{\theta}^2 - k(\t)\, \theta^2 \big] \ , \qquad \dot{\theta} \equiv \del_\t \theta \ ,
\la{orig}
\ee
one may redefine $\t$ as $\t \to t(\t)$, $\dot{t}(\t) = h^{-1}(\t)$ to put all time dependence in the harmonic potential term
\be
S = \int dt \, \big[ {\theta'}^2 - m^2(t)\, \theta^2 ] \ , \qquad \qquad m^2(t) = k(\t(t)) h(\t(t)) \ , \ \ \ \theta ' \equiv \del_t \theta \ .
\ee
The corresponding equation of motion is
\be
\theta'' +m^2(t)\, \theta = 0 \ . \la{eoml}
\ee
It is easy to see that for a given function $\theta_0(t)$, the quantity
\be
Q = \theta_0 \, \theta' - {{\theta}'_0} \, \theta \ ,
\ee
is conserved on-shell if and only if $ \theta_0$ is a particular solution of the equation of motion \rf{eoml}.
Furthermore,
such a conserved charge provides a first integral for \rf{eoml}
\be
Q = {\theta_0} \, \theta' - \theta'_0 \, \theta = C_1 = \text{const} \ \
\quad \to \quad (\tfrac{\theta}{\theta_0})' = \tfrac{C_1}{\theta_0^2} \ . \la{exact}
\ee
Integrating this first-order equation
yields the general solution of \rf{eoml} ($C_2=\const$)
\be\la{d6}
\theta(t) = C_1\, \theta_0(t) \int \tfrac{dt }{\theta_0^2(t)} +\ C_2\, \theta_0(t) \ ,
\ee
with $\theta=\theta_0$ being, of course, a special case.
Thus \rf{eoml} is solvable if a particular solution $\theta_0$ can be constructed explicitly.

Changing back to the original parametrization \rf{orig} ($t \to \t$), the conserved charge \rf{exact}
takes the form
\be
Q = h(\t) \big[ {\theta_0} \, \dot{\theta} - \dot{\theta}_0 \, \theta \big] \ , \la{lic}
\ee
where $\theta_0=\theta_0(\t)$ is a particular solution of \rf{orig}, while \rf{d6} becomes
\be
\theta(\t) = C_1\, \theta_0(\t) \int \tfrac{d\t}{h(\t) \, \theta_0^2(\t)} + C_2\, \theta_0(\t) \ . \la{gena}
\ee
The linearized theory \rf{lt} corresponds to \rf{orig} with $h(\t) = k(\t) = \t$.
The conserved charge in \rf{ql},\rf{ql2} is indeed of the form \rf{lic}.
From the monodromy matrix one finds that
$Q = \t [ \g(\t) \, \theta + \bw(\t) \, \dot{\theta} ]$ where
\be
\hspace{-0.2cm}\g(\t) = \int_{-\infty}^{+\infty} d\s \,e^{-im \s} \, \tfrac{m (s_+ s_- -w-\s)}{2 \t s_+ s_- } ,\ \bw(\t) = \int_{-\infty}^{+\infty} d\s\, \, e^{-im \s}\, \tfrac{ i}{2 s_+ s_- } , \ \ \ \ s_\pm \equiv \sqrt{w+\s\pm \t} \ . \la{ints}
\ee
Here $\g = - \dot \bw$ as required. Indeed, the term proportional to $\dot{\theta}$ in the leading $\O(\e)$
expansion of the flatness equation $\del_\t L_\s - \del_\s L_\t + [L_\t,L_\s] = 0$ tells us that the integrands in \rf{ints}
satisfy $
\del_\t [ \tfrac{e^{-im \s} i}{2 s_+ s_- } ] + \tfrac{e^{-im \s} m (s_+ s_- -w-\s)}{2 \t s_+ s_- } = \del_\s\, [\tfrac{e^{-im \s}\, i (s_+ s_- -w-\s)}{2 \t s_+ s_- } ] $.
Integrating over $\s$ and noting that $ \tfrac{e^{-im \s}\, i (s_+ s_- -w-\s)}{2 \t s_+ s_- } $ vanishes at $\s=\pm \infty$,\foot{This follows from the periodicity condition \rf{pe} since it is the coefficient of $\dot{\theta}$ in a component of the leading $\O(\e)$ term in $L_\t$.}
we conclude that $\gamma + \dot\bw=0$.

Since the charge in \rf{ql} is conserved on-shell (from the monodromy matrix construction),
it follows that $\bw(\t) $ is a particular solution, and thus the general solution \rf{gena} is in this case given by
\be\la{genn}
\theta = C_1\, \bw(\t) \int \tfrac{d\t}{\t\, \bw^2(\t)} + C_2\, \bw (\t)\ .
\ee

\section{Local currents in `chiral' theory}\la{chiral}
\def\theequation{D.\arabic{equation}}
\setcounter{equation}{0}

Apart from non-local conserved charges, integrable $\s$-models typically have
conserved charges associated to
local higher-spin currents (see, e.g., \cite{Goldschmidt:1980wq,Evans:1999mj,Evans:2000qx}).
Below we show that a class of
similar local currents exists in generalized $\s$-models where the couplings depend not
on $\t$ but on the l.c. variable $\xi^-=\ha (\t-\s)$,
\be
\widehat{\L}_{\rm ch} = G_{ij}( \xi^-, x)\, \del_+ x^i \del_- x^j \ . \la{e1} \ee
The classical action is then still invariant under `half' of the conformal transformations
$\xi^+\to f( \xi^+)$.
Note that such models can be obtained from the time-dependent models \rf{lcg} by rescaling $\xi^+ \to b \xi^+$ and taking the limit $b\to 0$.

In the ordinary (time-independent) integrable models in Table \ref{tab1}
one can construct special conserved higher-spin \textit{local} currents
as follows (we follow the notation in \rf{gl},\rf{sl})
\begin{align}
\del_\pm \mathcal{J}_\mp^{(n)} = 0 \ , \qquad \qquad &\mathcal{J}_\pm^{(n)} = \begin{cases}
d_{a_1 \cdots a_n} \mc{A}_\pm ^{a_1} \cdots \mc{A}_\pm^{a_n} \ & \text{(group} \ G) \ \la{cur}\\
d_{a_1 \cdots a_n} \mc{P}_\pm ^{a_1} \cdots \mc{P}_\pm^{a_n} \ & \text{(symmetric space}\ G/H) \
\end{cases} \\
&d_{a_1 \ldots a_n} = d_{(a_1 \ldots a_n)} \ , \qquad\qquad
{f^{a}}_{b(c} d_{a_1 \cdots a_{n-1}) a}= 0 \ .\la{e3}
\end{align}
The conservation of the currents \rf{cur} follows from the equations of motion
\begin{align}
G \ : \qquad &\del_+ \mc{A}_- + \del_- \mc{A}_+ = 0 \ , \qquad F_{+-}(\mc{A}) = 0 \ , \\
G/H \ : \qquad &D^\mc{B}_+ \mc{P}_- + D^\mc{B}_- \mc{P}_+ = 0 \ , \qquad F_{+-}(\mc{B} + \mc{P}) = 0 \ ,
\end{align}
since these may be re-written as
\begin{align}
G \ : \qquad &\del_\pm \mc{A}_\mp = \ha [A_\mp, A_\pm] \ , \la{ni}\\
G/H \ : \qquad &\del_\pm \mc{P}_\mp = [P_\mp, B_\pm ] \ , \ \ \ \qquad \ \ F_{+-}(\mc{B}) + [\mc{P}_+, \mc{P}_-] = 0 \ . \la{ni2}
\end{align}
Such higher-spin currents were
systematically studied in the group \cite{Evans:1999mj} and symmetric space \cite{Evans:2000qx} cases. One natural choice for the invariant
tensor $d_{a_1 \cdots a_n}$ is given by the symmetrized trace of the generators,
$d_{a_1 \cdots a_n} = \Tr[ T_{(a_1} \cdots T_{a_n)} ] $.

Suppose we now promote the corresponding
couplings to functions of the space-time coordinates, $h_\a \to h_\a(\t,\s)$, specially
chosen so that the resulting model still admits a Lax connection
as in \rf{glt},\rf{slt},\rf{laxs}.
It follows from the Lax representation that the equations of motion must now take the form
(cf.\ Appendix \ref{A})
\begin{align}
G \ : \qquad &\del_+ \mc{A}_- + \del_- \mc{A}_+ = a^\a \del_- h_\a \, \mc{A}_+ + b^\a \del_+ h_\a \, \mc{A}_- \ , \\
&\quad F_{+-}(\mc{A}) = c^\a \del_- h_\a \, \mc{A}_+ + d^\a \del_+ h_\a \, \mc{A}_- \ , \\
G/H \ : \qquad &D^\mc{B}_+ \mc{P}_- + D^\mc{B}_- \mc{P}_+ = a^\a \del_- h_\a \, \mc{P}_+ + b^\a \del_+ h_\a \, \mc{P}_- \ , \\
&\quad F_{+-}(\mc{B} + \mc{P}) = c^\a \del_- h_\a \, \mc{P}_+ + d^\a \del_+ h_\a \, \mc{P}_- \ ,
\end{align}
where $a^\a,b^\a,c^\a,d^\a$ are particular functions of $h_\a$ and $(\t,\s)$.
In general, these modified equations do not admit a form like \rf{ni},\rf{ni2}. However, in the `chiral' case where $h_\a = h_\a(\xi^-)$, we may set $b^\a=d^\a=0$ and then
it follows that
\begin{align}
G \ : \qquad &\del_- \mc{A}_+ = \ha [\mc{A_+, A_-}] + \ha(a^\a-c^\a) \del_- h_\a \, \mc{A}_+ \ , \la{mi}\\
G/H \ : \qquad &\del_- \mc{P}_+ = [\mc{P}_+, \mc{B}_- ] + \ha(a^\a-c^\a) \del_- h_\a \, \mc{P}_+\ . \la{mi2}
\end{align}
Thus, while the currents \rf{cur} are not conserved,
the `holomorphic' half of them $\mathcal{J}_+^{(n)} $ satisfy
\begin{align}
\del_- \mathcal{J}_+^{(n)} = \tfrac{n}{2} (a^\a-c^\a) \del_- h_\a \, \mathcal{J}_+^{(n)} \ .
\end{align}
This leads to the following modified holomorphic conservation law
\be\la{e15}
\del_- \widehat {\mathcal{J}}_+^{(n)} =0 \ , \qquad \qquad
\widehat {\mathcal{J}}_+^{(n)} \equiv
e^{- \tfrac{n}{2} \int dx^- (a^\a-c^\a) \del_- h_\a} \mathcal{J}_+^{(n)} \ .
\ee

\section{Lax pair in time-dependent sine-Gordon model}\la{sing}
\def\theequation{E.\arabic{equation}}
\setcounter{equation}{0}

As discussed in section \ref{s6}, the sine-Gordon model,
\be
\L = \tfrac{1}{g^2} \big[ \ha \del_+ x \del_- x + m^2 \cos{x}\big] \ , \la{sgma}
\ee
displays the same pattern as the $\s$-models considered above: upon promoting the couplings $(g,m)$ to functions of 2d time $\t$, the Lax connection naturally generalizes to the resulting time-dependent model only if the time dependence
is given by the 1-loop RG flow of the original model
\be m^2(\t) = e^{ \b(g)\, \t} m_0^2 \ , \ \ \ \qquad g(\t) = g \ , \ \ \ \ \qquad \b(g) = - 2 + g^2 \ . \la{52a} \ee
Below we shall justify this claim in more detail.

As was noted in section \ref{s6}, the time-dependent theory obtained from \rf{52a}
\be
\widehat{\L} = \tfrac{1}{g^2} \big[ \ha \del_+ x \del_- x + e^{\b(g)\, \t}\, m^2_0\, \cos{x}\big]\ , \la{53a}
\ee
is clearly integrable
since the explicit $\t$-dependence in \rf{53a} can
be removed by a 2d conformal transformation getting back to \rf{sgma}.
Indeed, starting with the Lax pair \cite{Lax:1968fm} for the original
sine-Gordon model \rf{sgma} ($\s_i$ are Pauli matrices)
\begin{align}
\hspace{-2cm} L_\pm = \pm \tfrac{i}{4} \del_\pm x \ \s_3 + \tfrac{i}{2} z^{\pm 1} \, m \cos{\tfrac{x}{2}} \ \s_1 \pm \tfrac{i}{2} z^{\pm 1} \, m \sin{\tfrac{x}{2}} \ \s_2 \ , \la{sgla}
\end{align}
and applying a conformal transformation $\xi^\pm \to f^\pm (\xi^\pm)$, one obtains a Lax connection
for the time-dependent theory \rf{53a},\foot{Note that, as for the $\s$-models discussed above, the dependence on the spectral parameter $z$ in \rf{larga} is again correlated with
a constant shift of $\s$.}
\be
\widehat{L}_\pm =\pm \tfrac{i}{4} \del_\pm x \ \s_3 + \tfrac{i}{2} z^{\pm 1} \, e^{\b(g)\, \xi^\pm} \, m_0 \cos{\tfrac{x}{2}} \ \s_1 \pm \tfrac{i}{2} z^{\pm 1} \,e^{\b(g) \, \xi^\pm} \, m_0 \sin{\tfrac{x}{2}} \ \s_2 \ . \la{larga}
\ee
Note that this Lax connection follows the same ansatz \rf{lac2} as in the \sm case:
it is obtained from the original Lax connection \rf{sgla} by the replacements
$z \to \widehat{z}= e^{\frac12 \b(g)\s}z$ and $m_0 \to m(\t) = e^{\frac12 \b(g)\t} m_0$.

Conversely, suppose we replace the couplings $(g,m)$ in \rf{sgma}
with general functions of $(\t,\s)$ and then demand that the resulting theory (cf.\ \rf{sth})
\be
\widehat{\L} = \tfrac{1}{g^2(\t,\s)} \big[ \ha \del_+ x \del_- x + m^2(\t,\s) \, \cos{x}\big] \la{54a}
\ee
admits a Lax representation. Motivated by \rf{sgla}, we shall assume the following ansatz for the Lax connection (cf.\ \rf{glt},\rf{slt})
\begin{align}
&\widehat{L}_\pm = f_{\pm}(\t,\s)\ \tfrac{i}{4} \del_\pm x \, \s_3 + \, v_\pm (\t,\s) \ \tfrac{i}{2}\cos{\tfrac{x}{2}} \, \s_1 + w_\pm (\t,\s) \ \tfrac{i}{2}\sin{\tfrac{x}{2}} \, \s_2\ . \la{sga}
\end{align}
Then
matching coefficients of various terms in the zero-curvature condition for \rf{sga} and in
the equation of motion corresponding to \rf{54a} leads to the following constraints on the coefficient functions in \rf{sga}
and the coupling functions,
\begin{align}
&f_\pm = \pm 1 \ , \qquad\qquad w_\pm = \pm v_\pm \ , \qquad\qquad \del_\mp v_\pm = 0 \ , \la{fe1} \\
&g(\t,\s) = g= \const \ , \qquad\qquad v_+ v_- = m^2(\t,\s) \ . \la{fe2}
\end{align}
It follows from \rf{fe1},\rf{fe2} that $v_\pm = \pm w_\pm = v_\pm(\xi^\pm)$, \ $m^2(\t,\s)=v_+(\xi^+)\, v_-(\xi^-)$.
Finally, applying a conformal transformation to \rf{sga},
we may set, e.g., $v_\pm(\xi^\pm) \to z^{\pm 1} \, e^{\b(g) \, \xi^\pm} \, m_0$, $m^2(\t,\s) \to e^{\b(g) \, \t} \, m_0^2$,
thus bringing the Lagrangian \rf{54a} and the Lax connection \rf{sga} to the form \rf{53a} and \rf{larga}
where the time dependence is given by the 1-loop RG flow \rf{52a} of the standard sine-Gordon model.

\small


\end{document}